\documentclass[aps,prb,twocolumn,longbibliography]{revtex4-2}

\usepackage{graphicx}
\usepackage{graphicx,epstopdf}
\usepackage{xcolor, soul}
\usepackage{dcolumn}
\usepackage{amsmath} 
\usepackage[colorlinks=true,linkcolor=blue,citecolor=blue]{hyperref}%
\usepackage{fontenc}
\usepackage{float}
\usepackage{amsthm}
\usepackage{subfigure}
\usepackage{color}
\usepackage{ragged2e}
\usepackage{bm}
\usepackage{enumerate}
\usepackage{hyperref}
\usepackage{amssymb}

\usepackage{xr}
\usepackage{comment}

\def\bea{\begin{eqnarray}}
\def\eea{\end{eqnarray}}
\def\be{\begin{equation}}
\def\ee{\end{equation}}
\def\no{\nonumber}
\def\v{\vec{q}}
\begin{document}

\title{Signature of quantum phase transition manifested in quantum fidelity \\
at finite temperature}

\author{Protyush Nandi} 

\email{protyush18@gmail.com}

\affiliation{Department of Physics, University of Calcutta,
92 Acharya Prafulla Chandra Road, Kolkata 700009,
India}

\author{Sirshendu Bhattacharyya}

\email{sirs.bh@gmail.com}

\affiliation{Department of Physics, Raja Rammohun Roy Mahavidyalaya,
Radhanagar, Hooghly 712406, India}

\author{Subinay Dasgupta}  

\email{subinaydasgupta@hri.res.in}

\affiliation{Department of Physics, Harish-Chandra Research Institute, Prayagraj 211019,
India}

\begin{abstract}
\noindent
The signature of quantum phase transition is generally wiped out at finite temperature. A few quantities that have been observed to carry this signature through a nonanalytic behavior are also limited to low temperatures only. With an aim to identify a suitable dynamical quantity at a high temperature, we have recently constructed a function from quantum fidelity, which has the potential to bear a nonanalytic signature at the quantum critical point beyond low temperature regime. 
In this paper, we elaborate our earlier work and demonstrate the behavior of the corresponding rate function and the robustness of the nonanalyticity for a number of many-body Hamiltonians in different dimensions. We have also shown that our rate function reduces to that used in the demonstration of the dynamical quantum phase transition (DQPT) at zero temperature. It has been further observed that, unlike DQPT, the long time limit of the rate function can faithfully detect the equilibrium quantum phase transition as well.
\end{abstract}

\maketitle

\section{Introduction}
\noindent
Quantum many-body systems, when brought to nonzero temperatures, give rise to interesting dynamical features. In the low temperature regime, both the quantum and the thermal fluctuations come into play within a quantum critical region. However, the quantum counterpart recedes when we move towards higher temperatures. Therefore the detection of the phenomena driven by quantum fluctuations, for instance the quantum criticality is generally limited to low temperatures only \citep{sachdev_2011,hofferberth2008,campisi2011,
greiner2002,haller2010,zhang2012,guan2013}. 
On the other hand, such detection opens up a new challenge for tracing the imprint of quantum critical phenomena at finite temperatures \citep{sondhi1997}.

\par Quantum fidelity has been considered to be a powerful tool in the detection of the quantum phase transition (QPT). Defined as a measure of the overlap between the eigenstates of the pre-quench and post-quench Hamiltonians, fidelity can ideally vary from $1$ to $0$. For a quench across quantum critical point, this value comes close to zero showing a sharp dip because of the structurally different ground states of two phases \citep{zanardi2006,zanardi2007jsm,cozzini2007,zhou2008,gu2010,damski2016}. Different forms of fidelity has been studied at finite temperatures as well \citep{jacobson2011,zanardi2007,amin2018,quan2009,dai2017,liang2019,michal2021,mera2018} and some of them have been able to detect the quantum critical point by showing  nonanalyticity in their logarithms at low temperatures  \citep{jacobson2011,zanardi2007}.

\par As the QPT cannot be observed beyond zero temperature, lots of investigations have been made in search of its trace in different response functions. Quantum coherence has been one of the important functional candidates \cite{Li_new}. At low temperatures, the absorbed energy is reported to be able to detect the quantum critical region \cite{haldar2020}. Another related work shows that the expectation values of local order parameters at infinite time exhibiting nonanalytic behavior at critical points at low temperatures \cite{roy2017}. However it vanishes as the temperature increases. Recently a form of rate function for Loschmidt amplitude studied at finite temperature, has been shown to detect the presence of QPT \cite{Hou222}. The time dependences of the work distribution function, or the magnetization following quantum quenches or pulses across quantum critical points also bear signatures of QPT \cite{abeling2016, zvyagin2018, zvyagin_pulse, zvyagin_noneq, zvyagin_review, zvyagin_book2010}. Another quantity, the out of time ordered correlations (OTOC) is found to detect quantum critical point through its time averaged form at infinite temperature for 1D ANNNI model {\cite{souvik_otoc}}. Quantum teleportation protocol applied on 1D XXZ model can also signal quantum critical point at finite temperatures \cite{Rigolin_teleport,Ribeiro2023Oct}. 

The above findings motivated us to look for a quantity which will be able to bear signatures of QPT at appreciable temperatures in a $d$-dimensional system. In an earlier work, we had defined a form of fidelity in which we could trace down a nonanalytic signature in its rate function at the QCP (quantum critical point) even at high temperatures \citep{nandi2022}.  At zero temperature, the fidelity reduces to Loschmidt echo which is used to define dynamical quantum phase transition \cite{Heylprl,Karrasch2013,Kriel2014,Canovi2014,Heyl2014,Hickey2014,Andraschko2014,James2015,Vajna2015Apr,Heyl2015oct,Budich2016feb,abeling2016,Vajnaprb}. Here in this paper, we extend our earlier work and report a comprehensive extended study on the response of the detector in 1D, 2D and 3D systems.  We confined ourselves to the search for dynamical characteristics at high temperatures. 

\par In 3-dimension, the topological Weyl semimetals have recently emerged as a novel material with unique electronic structures. Here the electrons effectively behave as relativistic Weyl fermions when the Fermi energy is near the crossing points of valence and conduction bands. Unlike topological insulators or topological Dirac semimetals, the Weyl semimetals have broken time reversal symmetry which leads to a topological phase transition. At zero temperature, Weyl semimetals exhibit a gapped to gapless phase transition. A similar phase transition also occurs in topological nodal line semimetals as well. The only difference between the two is that the Hamiltonian of the Weyl semimetals consists of isolated gapless points in the fermionic momentum space whereas in case of the nodal line semimetals, they form gapless lines. We demonstrate numerically that the rate function mentioned above shows nonanalyticity right at the critical points in both the cases even at finite temperatures. 

\par  We also apply our methodology to the quantum spin Hall insulator Bi$_4$Br$_4$ using the $k\cdot p$ Hamiltonian proposed for this material
 \cite{Yao2015, Hasan2022}
and find that our (numerically computed) rate function predicts the location of quantum phase transition correctly for this material at low and high temperature. 
 
\par This paper is organized as follows. In Section \ref{Theory} we provide the definition of the functional form of the fidelity along with the rate function and derive the expressions for a general $d$-dimensional system. We study the rate function for one dimensional XY model in Section \ref{1D} and for the 1D Su-Schrieffer–Heeger (SSH) model in Section \ref{1d_ssh}. Section \ref{2D} contains the same for the 2D Kitaev model on a honeycomb lattice. The results for 3D systems namely, Weyl semimetal, topological nodal line semimetal and the material Bi$_4$Br$_4$ are presented in Section \ref{3D} . In Section \ref{t0case}, we discuss the behaviour of the detector at zero temperature. This is followed by concluding remarks in Section \ref{conclu}. 

\section{Theory} \label{Theory}
Let us consider a system with Hamiltonian $\mathcal{H}$, dependent on a parameter which will be quenched from $p_0$ to $p$ at time $t=0$ so that the Hamiltonian is quenched from ${\mathcal{H}}_0=\mathcal{H}(p_0)$ to $\mathcal{H}'=\mathcal{H}(p)$. We assume that the system was initially in thermal equilibrium at inverse temperature $\beta$.  We now define quantum fidelity as
\bea \mathcal{F}_t \equiv \frac{ {\rm Tr}[ \rho_t \cdot \rho_0]}  {{\rm Tr} \left[ \rho_t \right]\, {\rm Tr} \left[ \rho_0 \right]} \label{def-Fd} \eea
where $\rho_0$ is the density matrix at $t=0$ and $\rho_t$ is the same after the system has evolved through time $t$.
\bea \rho_0 &=& \exp{(-\beta {\mathcal{H}}_0)} \nonumber\\ \rho_t&=& \exp{(-i\mathcal{H}'t)}\exp{(-\beta {\mathcal{H}}_0)}\exp{(i\mathcal{H}'t)} \eea 
 One may note that the probability of overlap between initial and time-evolved states namely $|\langle \psi(0) |\psi(t) \rangle|^2$, is known as the Loschmidt echo which shows nonanalytic kinks in its logarithm  as a function of time when the Hamiltonian is quenched across a QCP \citep{Heyl_2018}. Our expression for fidelity reduces to this Loschmidt echo, when brought to zero temperature.
We may define a measurable quantity called rate function as,
\bea r(t,\beta,p_0,p) \equiv  - \lim_{N\to\infty}\frac{1}{N} \log {\mathcal F}_t \label{r} \eea
\noindent
 where $N$ is the system size. The quantity of our interest  is the long-time average of this rate function, defined as
\bea r_a(\beta,p_0,p) \equiv \lim_{\tau \to \infty} \frac{1}{\tau} \int_0^\tau  r(t,\beta,p_0,p) \, dt \label{r-inf}\eea
This is the detector we shall use to locate the presence of QCP at a finite temperature. We shall investigate if the behaviour of this quantity, at zero and nonzero temperatures bears a signature of QCP. 
In order to study this rate function we consider a class of systems with a generic Hamiltonian expressible as a sum of commuting $2\times2$ Hamiltonians in the space of wave-vectors $\vec{q}$
\bea \mathcal{H}=\sum_{\vec{q}} \mathcal{H}_{\vec{q}} \label{genf}\eea
\noindent
and express $\mathcal{H}_{\vec{q}}$ as
\be \mathcal{H}_{\vec{q}}=a_{\vec{q}}\sigma_1 +b_{\vec{q}}\sigma_2+c_{\vec{q}}\sigma_3 = \lambda_{\vec{q}}(\hat{\mathcal{V}}_{\vec{q}}\cdot \vec{\sigma}) \label{expression_Hq} \ee
Here $\sigma_i$ are the Pauli spin matrices, $\lambda_{\vec{q}}=\sqrt{a_{\vec{q}}^2+b_{\vec{q}}^2+c_{\vec{q}}^2}$, $\hat{\mathcal{V}}_{\vec{q}}=\left({a_{\vec{q}}}/{\lambda_{\vec{q}}},{b_{\vec{q}}}/{\lambda_{\vec{q}}},{c_{\vec{q}}}/{\lambda_{\vec{q}}}\right)$ where $\v$ is $d$-dimensional wave vector. After quenching $p_0\rightarrow p$, the Hamiltonian becomes
$${\mathcal{H}}'_{\vec{q}}={\mathcal{H}}_{\v} (p) =\lambda'_{\vec{q}}({\mathcal{V}}'_{\vec{q}}\cdot \vec{\sigma}) $$ 
Here ${\lambda'}_{\vec{q}} = \sqrt{{a'}_{\vec{q}}^2+{b'}_{\vec{q}}^2+{c'}_{\vec{q}}^2}$, $\hat{\mathcal{V}'}_{\vec{q}} = \left({{a'}_{\vec{q}}}/{{\lambda'}_{\vec{q}}},{{b'}_{\vec{q}}}/{{\lambda'}_{\vec{q}}},{{c'}_{\vec{q}}}/{{\lambda'}_{\vec{q}}}\right)$ and the primed quantities are the post quench values. We can calculate the exponentials in $\rho_0$ and $\rho_t$ by exploiting the property that 
\[ \mathcal{H}_{\vec{q}}^2=\lambda_{\vec{q}}^2 \underline{1} \;\;\;
\mbox{and} \;\;\; {\mathcal{H}'_{\vec{q}}}^2={\lambda'_{\vec{q}}}^2 \underline{1}\]

$(\hat{\mathcal{V}}_{\vec{q}}\cdot \vec{\sigma} )^2=1$ and obtain the fidelity in Eq. [\ref{def-Fd}] as
\bea {\mathcal{F}_t}(t,\beta,p_0,p)= \frac{1}{2} \left( 1+\tanh^2(\beta	\lambda_{\v})[1- \right. \no \\ \left. 2\sin^2(\lambda' _{\v} t)\{1-(\hat{{\mathcal{V}}'}_{\vec{q}}.\hat{{\mathcal{V}}}_{\vec{q}})^2\}] \right)  \eea 
Hence the rate function (\ref{r}) of the generic Hamiltonian (\ref{genf}) can be written as 
 \bea r(t,\beta,p_0,p) = \log 2 - {\frac{1}{V}}\int_{\vec{q}} \; d{\vec{q}} \; \log \left[ 1+ \right.\nonumber \\
\left. \tanh^2 (\beta \lambda_{\vec{q}}) \left\{1 - 2\sin^2( \lambda'_{\vec{q}}  t) \, \mathcal{L}_{\vec{q}}\right\}
\right] \label{prev_eqn8} \eea  
in $d$ dimensions where the volume of the $d$-dimensional first Brillouin zone is written as $V=\int_{\vec{q}} d\vec{q}$, $\mathcal{L}_{\vec{q}}= 1 - (\hat{\mathcal{V}}_{\vec{q}}\cdot \hat{\mathcal{V}'}_{\vec{q}})^2=\sin^2(\phi_{\vec{q}})$, $\phi_{\vec{q}}$ is the angle between $\hat{\mathcal{V}}_{\vec{q}}$ and $\hat{\mathcal{V}'}_{\vec{q}}$. 
We can calculate the long time average of the rate function from the Eq. (\ref{r-inf}) using standard results \citep{gradshteyn2014}. 
\bea r_a(\beta,p_0,p) &=& 3\log 2-\frac{1}{V}\int_{\vec{q}} \log(1+\alpha_{\vec{q}}) d\vec{q}\nonumber \\ &-&\frac{2}{V} \int_{\vec{q}} \log \left[1+\sqrt{1-\gamma_{\vec{q}} \mathcal{L}_{\vec{q}}}\right] d\vec{q} \label{rddim} \eea 
where
 $\alpha_{\vec{q}}=\tanh^2(\beta \lambda_{\vec{q}})$, $\gamma_{\vec{q}}=2\alpha_{\vec{q}}/(1+\alpha_{\vec{q}})=1 - \text{sech} (2\beta \lambda_{\vec{q}})$.

Since both $\gamma_{\vec{q}}$ and $\mathcal{L}_{\vec{q}}$ have values between $0$ and $1$, we can expand the integrand in the third term of Eq.~(\ref{rddim}).
\bea r_a(\beta,p_0,p) &=& \log 2 - \frac{1}{V}\int_{\vec{q}}\log(1+\alpha_{\vec{q}}) d\vec{q} \nonumber \\ &+& \frac{1}{2V} \sum_{n=1}^{\infty} {c_n}\int_{\vec{q}}\gamma^n_{\vec{q}}{\mathcal{L}}^n_{\vec{q}} d{\vec{q}}\label{rlog1}\eea  
where $c_n$ are constants generated from the expansion and $c_1=1$, $c_2=\frac{3}{8}$, $c_3=\frac{5}{24}$ etc. For detailed derivation of Eqs. (\ref{prev_eqn8}), (\ref{rddim}) and (\ref{rlog1}), see Appendix~A.  
Our objective is to study the quantity ${r_a}(\beta,p_0,p)$ as a function of the post-quench value $p$ of this parameter. Actually we shall look for nonanalytic behaviour of the quantities $\partial {r_a}(\beta,p_0,p)/\partial p$ and/or $\partial^2 {r_a}(\beta,p_0,p)/\partial p^2$. While differentiating Eq (\ref{rlog1}), the first two terms become zero as they are not functions of $p$. Therefore the third term is our subject of interest.
For all the Hamiltonians considered below, we have verified numerically that the predominant contribution to the nonanalytic behaviour (if any) in the double derivative comes from the point (called node) where $\lambda'_{\vec{q}}$ becomes zero. 
Therefore, in the second integral in Eq. (\ref{rlog1}), $\lambda_{\vec{q}}$ and $\gamma_{\vec{q}}$ are brought out of the integral by replacing $\lambda_{\vec{q}}$ by its value at the node. So we can write 
\bea r_a(\beta,p_0,p) &=& \log 2 - \frac{1}{V}\int \log(1+\alpha_{\vec{q}}) d\vec{q} \nonumber \\ 
&+& \frac{1}{2V} \sum_n {c_n} \frac{\gamma^n} {\lambda^{2n}} I_n \label{rlog}\eea  
where $\lambda$ and $\gamma$ are the value of $\lambda_{\vec{q}}$ and $\gamma_{\vec{q}}$ at the node, and
 $$ I_n(p_0,p) = \int_{\v} {\lambda_{\v}}^{2n} {{\mathcal{L}}_{\v}}^{n}\; d\v $$
Following our earlier work \cite{nandi2022}, we shall show below that the nonanalyticity comes from the integral $I_n$ and since $I_n$ is independent of temperature, the signature of the existence of QCP is expected to show up at all temperatures. 

We shall apply the above prescription to several integrable quantum spin models, namely the XY chain, 1D SSH Model, the Kitaev model on a honeycomb lattice, Weyl semimetals and Topological Nodal Line Semimetals. Each of these models
shows a QCP at absolute zero temperature.
We shall show that the quantity $r_a$ shows a nonanalytic behaviour at the QCP at {\em any} finite temperature. Indeed, it does not imply that there is {\em actually} a phase transition at a finite temperature but that the detector, namely, the long-time fidelity, bears a signature of criticality which exists at zero temperature. Furthermore, since the evaluation of the relevant quantity only involves the evaluation of an integral, we can study numerically the 3-dimensional systems also. We shall report that for Weyl semimetals and Topological Nodal Line Semimetals, at low temperatures the long-time fidelity shows nonanalytic behaviour at the phase boundary. 

\section{Transverse Field XY Model}\label{1D}
\noindent
The XY model in 1D is defined by,
\bea {\mathcal H}_{XY} &=& - \frac{1}{2} (1+ h) \sum_{i=1}^N s_i^x s_{i+1}^x\\ \nonumber &-& \frac{1}{2} (1- h) \sum_{i=1}^N s_i^y s_{i+1}^y  
 -\Gamma  \sum_{i=1}^N s_i^z \label{XY-def1} \eea 
where $h$ is the anisotropy parameter and $\Gamma$ is the transverse field and $s_i$ are Pauli matrices. One can show by using Jordan-Wigner transformation \citep{sachdev_2011,LSM,pfeuty} that this Hamiltonian can be written in the form of Eq.~(\ref{genf}), where
\be {\mathcal H}_q = a_q\sigma_1 + b_q \sigma_3  \label{XY_def2}      \ee
Here $a_q = -h \sin (q)$ and $ b_q = \Gamma + \cos (q)$ with $0<q<\pi$. There is  a disordered phase in the region $|\Gamma| >1$ and two ordered phases for $h>0, |\Gamma| < 1$ and $h < 0, |\Gamma| < 1$.

 The long-time average of the rate function is now given by Eq. (\ref{rddim}) with $\vec{q}$ as a scalar.
We will study two quench protocols, one for the quench of external field ($\Gamma$) and the other for quench of the anisotropy parameter($h$). 


\subsection{Quench of External Field}

\begin{figure*}
\includegraphics[scale=0.42]{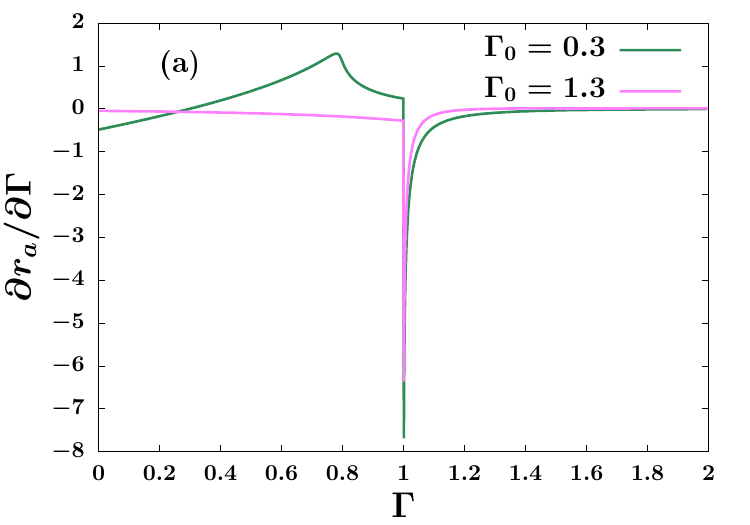}
\includegraphics[scale=0.42]{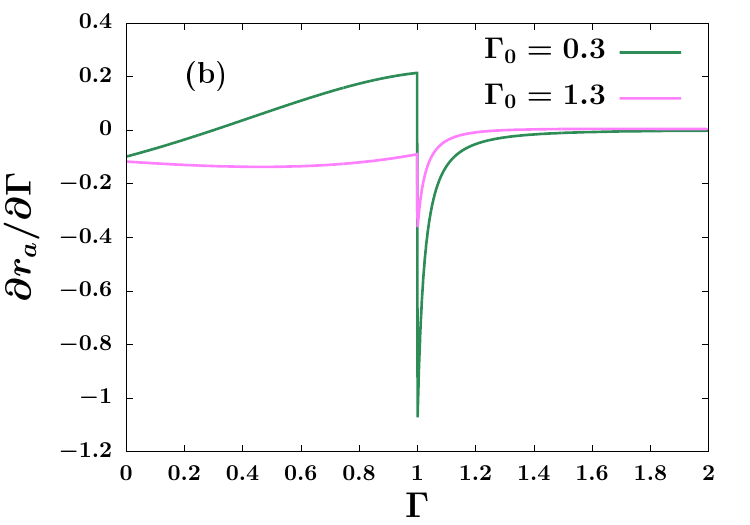}
\includegraphics[scale=0.42]{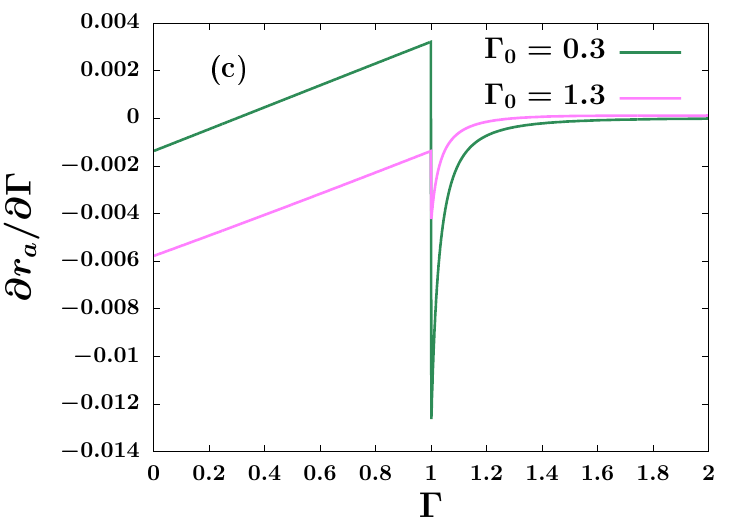}
\includegraphics[scale=0.42]{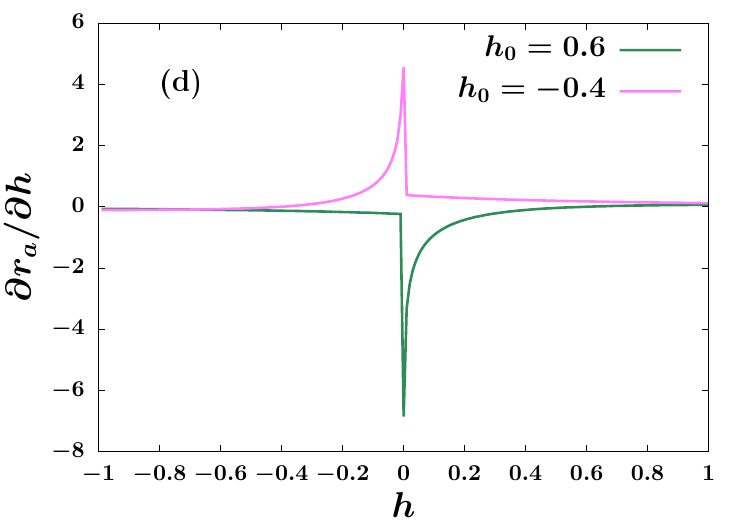}
\includegraphics[scale=0.42]{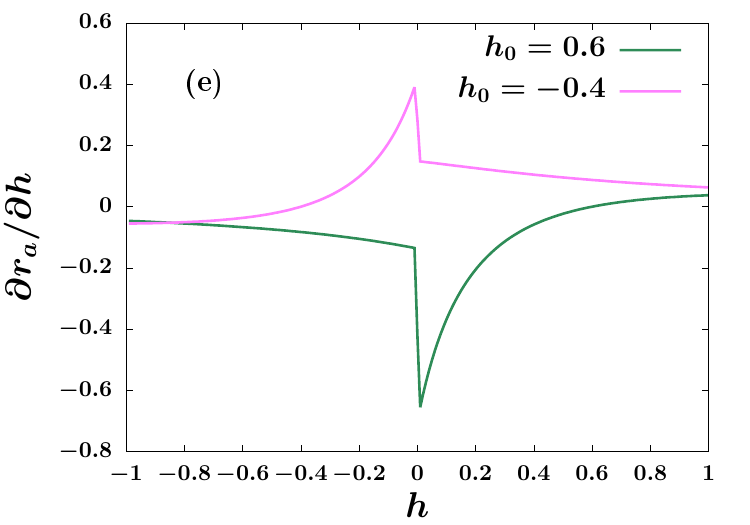}
\includegraphics[scale=0.42]{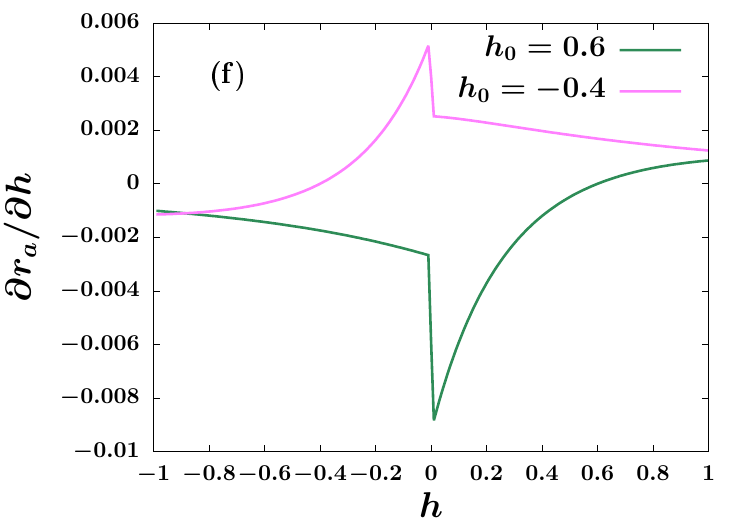}
\caption{ Transverse XY chain: First derivative of rate function with respect to the quench parameter, computed numerically  for quench of transverse field parameter $\Gamma$ for  $h=0.3$ at (a) $\beta=10$ ,(b) $\beta =1$ and (c) $\beta=0.1$ and for quench of anisotropy parameter $h$ for $\Gamma=0.3$  at (d) $\beta=10$ ,(e) $\beta =1$ and (f) $\beta=0.1$ In both cases the quantity $\partial r_a/\partial \Gamma$ is discontinuous at the critical point $\Gamma=1$ and $h=0$. }
\label{XY_Resuts}
\end{figure*}

We quench $\Gamma$ instantaneously from some initial value $\Gamma_0$ to $\Gamma$ keeping $h$ constant. The derivative $\partial r_a/\partial \Gamma$, calculated numerically shows 
  a discontinuity at $\Gamma=1$ which is the quantum critical point at the temperature $T=0$ (Fig. \ref{XY_Resuts}). This proves that our detector $r_a$ can successfully detect the QCP even at a large temperature as we have reported in our previous paper.\cite{nandi2022}

We can also derive an analytic expression for the amount of discontinuity at high temperatures from Eq. (\ref{rddim}). The parameter $p$ becomes $\Gamma$ in this case, $V$ becomes $\pi$ as the system is one dimensional and $q$ ranges from $0$ to $\pi$. The quantity $\mathcal{L}_q$ becomes \bea
\mathcal{L}_q = (\Gamma-\Gamma_0)^2 \left(\frac{a_q}{\lambda_q \lambda'_q}\right)^2 \no
\eea  Following Eq. (\ref{rlog}) we express $I_n$ , as
\bea I_n={(\Gamma-\Gamma_0)^{2n}}\int_{0}^{\pi} \left(\frac{a_q}{\lambda'_q}\right)^{2n} dq \eea
 For $n=1$, one can calculate the integral directly:  
\begin{eqnarray}
I_1=\frac{\pi}{4}{(\Gamma - \Gamma_0)^2}\left(1+\frac{1}{\Gamma^4}\right)\;\;\;\; {\mbox{for}} \; \Gamma >1
\end{eqnarray}
and
\begin{eqnarray}
I_1=\frac{\pi}{4}{(\Gamma - \Gamma_0)^2}\left(\Gamma^2+\frac{1}{\Gamma^2}\right) \;\;\;\; {\mbox{for}} \; \Gamma<1
\end{eqnarray} 
 When the temperature is high, $\beta$ is small and $I_1$ is the most dominant term(Refer to Appendix~C). 
The amount of discontinuity at the QCP $\Gamma=1$ is then
\bea (\partial r_a /\partial \Gamma)_{\Gamma=1+} - (\partial r_a /\partial \Gamma)_{\Gamma=1-} = \beta^2 (1-\Gamma_0^2)/h \eea
At high temperatures, we have calculated numerically the discontinuity in $\partial r_a /\partial \Gamma$ keeping all values of $n$, and found the result to be of the same order of magnitude as that obtained by keeping the $n=1$ term only.(Refer to Appendix~C)  

\subsection{Quench of Anisotropy Parameter}
In this case, we quench $h$ instantaneously from some initial value $h_0$ to $h$ keeping $\Gamma$ constant. The derivative $\partial r_a/\partial h$, calculated numerically shows 
a discontinuity at the phase boundary $h=0$ (Fig. \ref{XY_Resuts}). Thus, here also our detector $r_a$ can successfully detect the QCP even at a large temperature as we have reported in our previous paper\cite{nandi2022}. 

 The parameter $p$ of Eq (\ref{rddim}) becomes $h$ in this case and $V$ becomes $\pi$ for the reason mentioned in the previous subsection. $\mathcal{L}_q$ then becomes \bea
\mathcal{L}_q = (h-h_0)^2 \left(\frac{b'_q \sin (q)}{\lambda'_q \lambda_q}\right)^2 \no
\eea  We write
 the integral in the third term of Eq. (\ref{rlog}) as 
\bea I_n &=& (h-h_0)^{2n} \int \left(\frac{b'_{{q}} \sin (q)}{\lambda'_{{q}}}\right)^{2n} dq \label{xyanis2}\eea
 The term $I_1$ can be obtained as 
\begin{eqnarray}
I_1 &=& -\frac{\pi}{4}(h-h_0)^2 \{6h^2(2\Gamma^2- 1) \nonumber \\ &+&4h(\Gamma^2 -1) -2\}\;\;\;{\mbox{for}} \; h<0 \label{xyI1_1}
\end{eqnarray}
and
\begin{eqnarray}
I_1&=&-\frac{\pi}{4}(h-h_0)^2 \{2h^2(4\Gamma^2- 1) \nonumber \\&-&4h(\Gamma^2 -1) -2\}\;\;\; {\mbox{for}} \; h>0 \label{xyI1_2}
\end{eqnarray}
This gives the amount of discontinuity at the QCP $h=0$ at high temperatures as 
\bea (\partial r_a /\partial h)_{h=0+} - (\partial r_a /\partial h)_{h=0-} = 2\beta^2 h_0^2 (1-\Gamma^2) \label{xyanis_discont}\eea 
As before, the discontinuity calculated by keeping all values of $n$ turns out to be of the same order of magnitude as that obtained by keeping the $n=1$ term only (Refer to Appendix~C). 

\section{1D SSH Model}\label{1d_ssh}
\noindent
The Su-Schrieffer-Heeger (SSH) model can be thought of as a chain of $N$ unit cells with each unit cell consisting of two different sites labelled as $c$ and $d$ \citep{ssh1979,ssh1980}. The Hamiltonian in position space can be written as 
\bea \mathcal{H} = -\sum_{n} [c^{\dagger}_{n} d_{n} + h c^{\dagger}_{n+1} d_{n} + h.c]   \eea
where $c_{n}$ and $d_{n}$ are the fermionic annihilation operators at site $c$ and site $d$ respectively at the $n$-th unit cell and $c^{\dagger}$, $d^{\dagger}$ are the corresponding creation operators. 
After Fourier transformation, the Hamiltonian is transformed into the form of Eq.~(\ref{genf})
\bea \mathcal{H} = \sum_q \mathcal{H}_q\;\; \mbox{where} \;\; \mathcal{H}_q = a_q \sigma_1 +b_q \sigma_2  \eea
with $a_q = (1+h \cos (q))$ and $b_q = h\sin (q)$ where $q$ ranges from $-\pi$ to $\pi$. This  model shows phase transition at $h=\pm 1$.

\begin{figure*}
    \centering
    \includegraphics[scale=0.23]{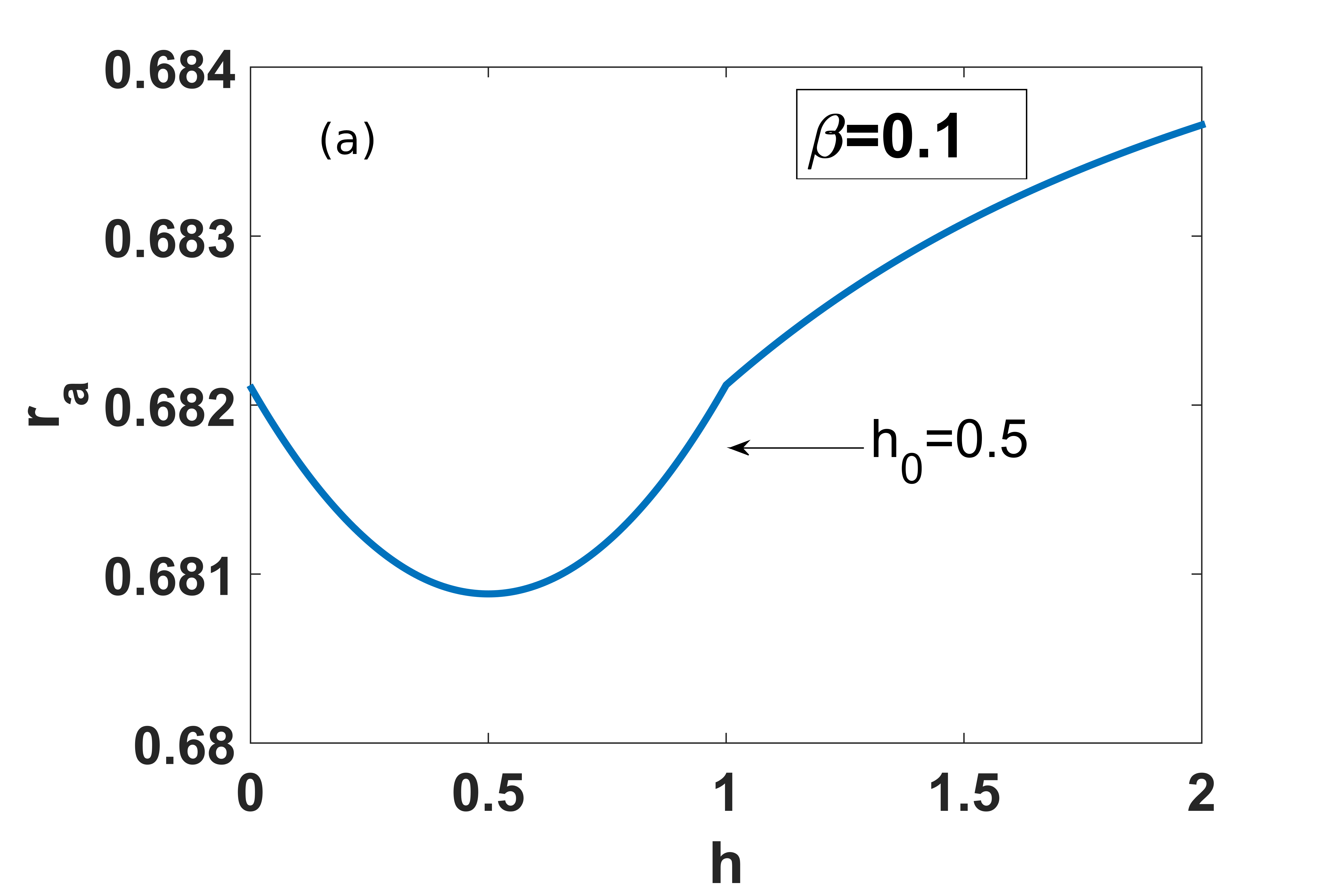}
    \includegraphics[scale=0.23]{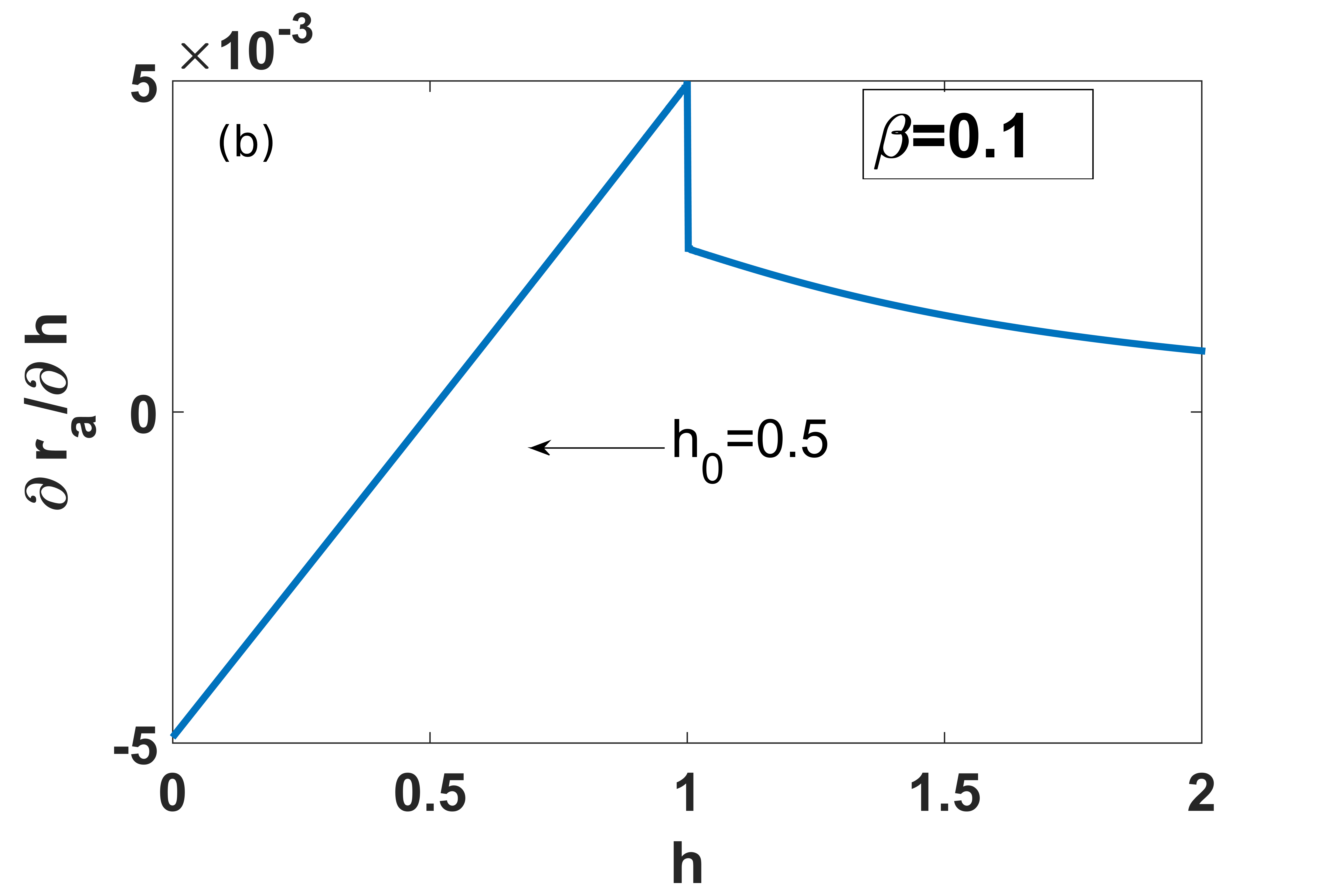}
    \includegraphics[scale=0.23]{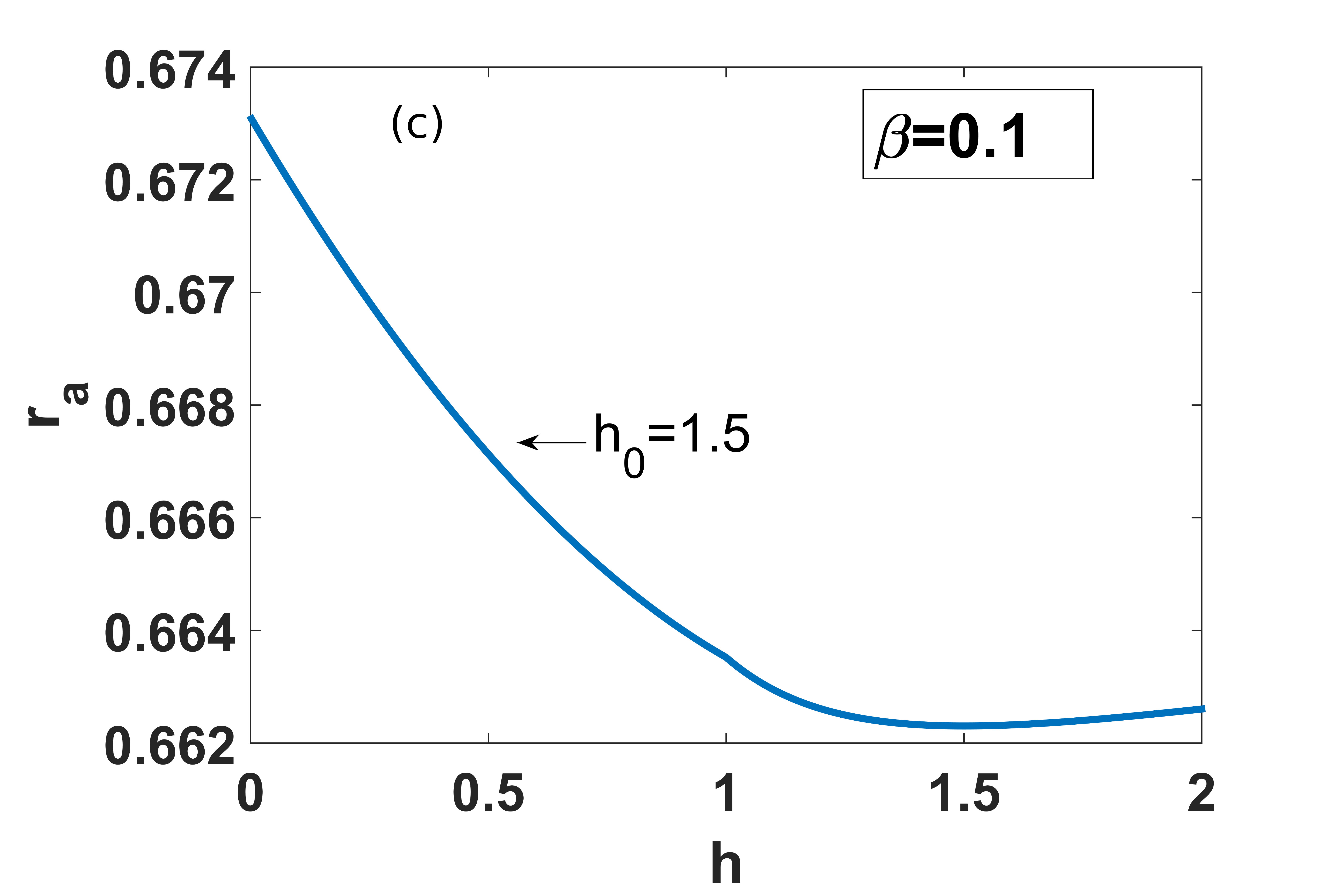}
    \includegraphics[scale=0.23]{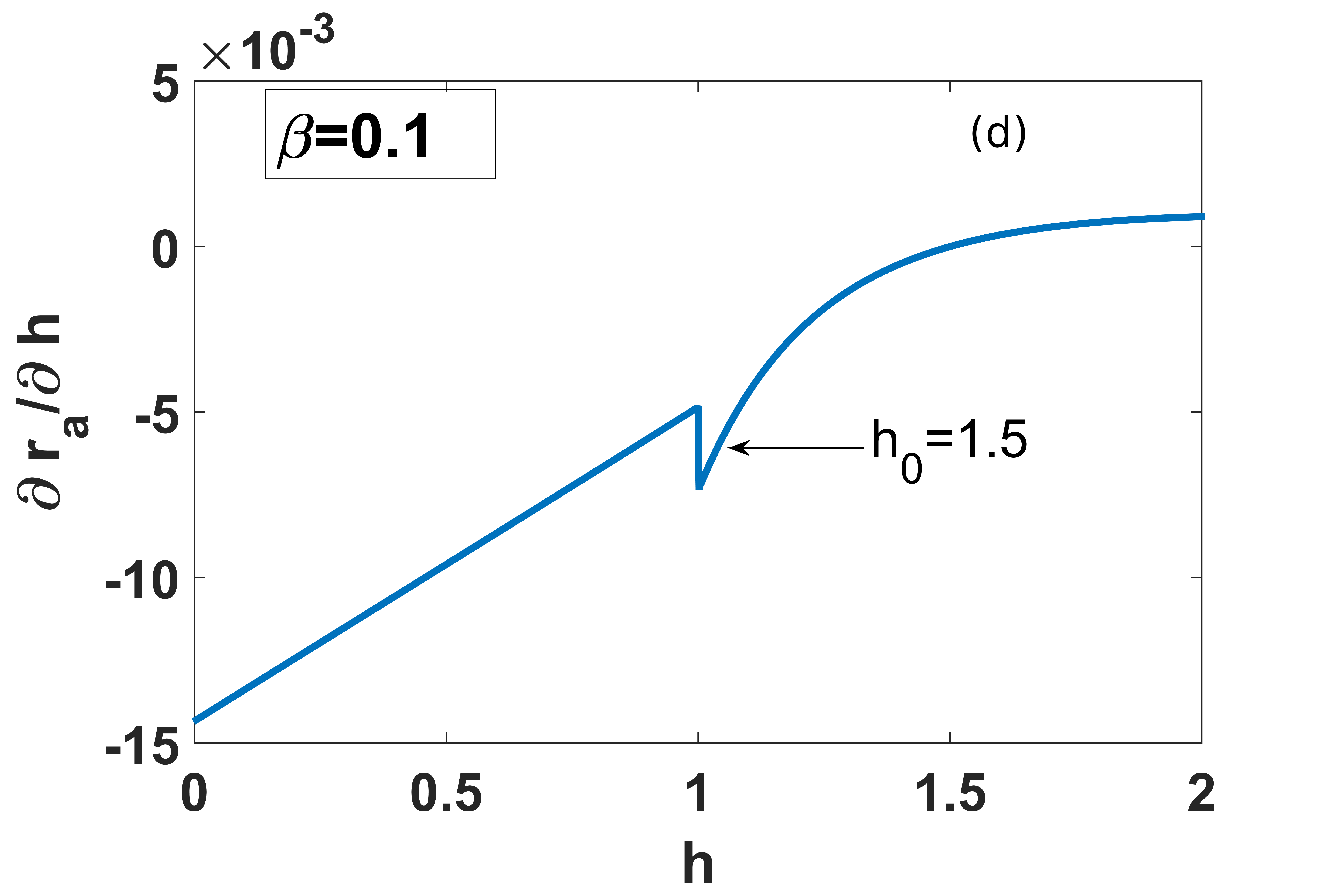}
    \caption{1D SSH model: The plots are the rate function and derivative of rate function with respect to post quench parameter $h$ for $h_0=0.5$ in (a) and (b) and $h_0=1.5$ in (c) and (d) at inverse temperature $\beta=0.1$ }
    \label{sshplot}
\end{figure*}

The parameter $h$ is quenched from $h_0$ to $h$ and the system is allowed to evolve. Using Eq. (\ref{rddim}), one can calculate numerically the rate function $r_a$ and its derivatives $\partial r_a/ \partial h$. One finds [Fig. \ref{sshplot}] a discontinuity in $\partial r_a/\partial h$ at $h=1$ which shows that the $r_a$ detects the QCP even at finite temperatures. We can analytically calculate the amount of discontinuity in $\partial r_a/\partial h$ at the QCP. 
The quantity $p$ becomes $h$ in this case, $V$ is equal to $2\pi$ and $\mathcal{L}_q$ is 
\bea \mathcal{L}_q = \left(\frac{a_{{q}}b'_{{q}}-b_{{q}}a'_{{q}}}{\lambda_{{q}} \lambda'_{{q}}}\right)^2 \eea
with the primed quantities being the post quench parameters.\\ 
We can now write the integral in the the third term of Eq. (\ref{rlog}) as
\bea I_n = \int_{q=-\pi}^{\pi}  {\left(\frac{a_{{q}}b'_{{q}}-b_{{q}}a_{{q}}'}{\lambda'_q}\right)}^{2n} dq \nonumber \\
=(h-h_0)^{2n}\int_{q=-\pi}^{\pi} \left(\frac{\sin^2 (q)}{1+h^2+2h\sin (q)}\right)^n dq 
\eea

At high temperature, the most dominant contribution comes from $I_1$, which can be calculated easily. 
\bea I_1 = \pi (h-h_0)^2 \;\;\; \mbox{for} \; h<1 \eea
and 
\bea I_1 = \pi\frac{(h-h_0)^2}{h^2} \;\;\; \mbox{for} \; h>1\eea
Hence, the amount of discontinuity in the first derivative of $r_a$ at $h=1$ (at high temperature) is,
\bea {\partial r_a}/{\partial h}|_{h=1+}-{\partial r_a}/{\partial h}|_{h=1-} = -2\beta^2(h_0-1)^2 \eea
 As before, the discontinuity calculated by keeping all values of $n$ turns out to be of the same order of magnitude as that obtained by keeping the $n=1$ term only (Refer to Appendix~C).

\section{Two Dimensional Case : Kitaev Model}\label{2D}
\begin{figure}[h!]
    \centering
    \includegraphics[scale=0.5]{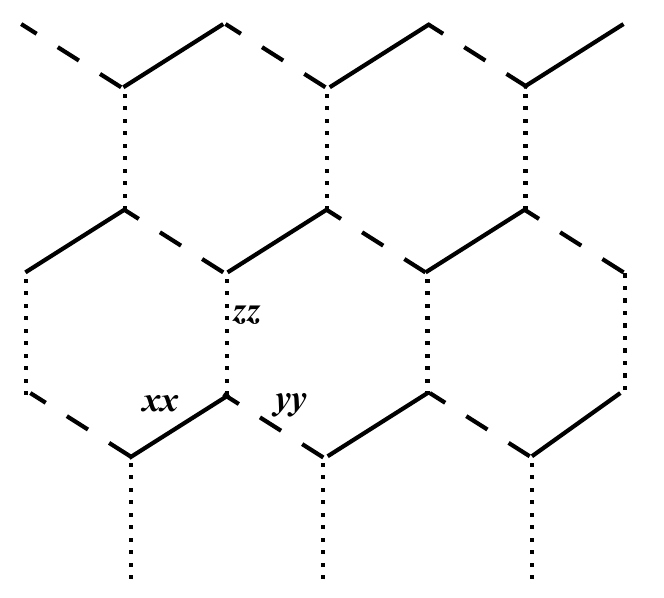}
    \caption{{Kitaev model on honeycomb lattice. The continuous, dashed and dotted lines correspond to $xx$, $yy$, and $zz$ interactions, with interaction strengths $J_1$, $J_2$, $J_3$ respectively. }}
    \label{kitaev_defn}
\end{figure}

The Hamiltonian of the Kitaev model on a honeycomb lattice is defined as,
\bea \mathcal{H} = \sum_{i,j} \sum_{\alpha=1}^3 J_{\alpha} \sigma_i^{\alpha} \sigma_j^{\alpha} \label{H_def1} \eea
where $i$, $j$ run over all the nearest-neighbouring pairs on the lattice. 
This model contains three interaction parameters $J_{1,2,3}$ (Fig.~\ref{kitaev_defn}). It can be shown \citep{kitaev2006,feng2007,kitaev2010} that in the ``vortex-free'' sector this Hamiltonian can again be written like Eq.~(\ref{genf}) with
\be {\mathcal H}_{\vec{q}} = a_{\vec{q}} \sigma_1 + b_{\vec{q}} \sigma_3  \label{H_def2} \ee
Here ${\vec{q}}=(q_x, q_y)$ where $-\pi<(q_x,q_y)<\pi$ and the coefficients are \citep{sengupta2008,senguptaPRB2008}
\bea 
a_{\vec{q}} &=& - J_1 \sin (q_x) +  J_2 \sin (q_y) \label{ak-bk} \nonumber \\
 b_{\vec{q}} &=& J_3 -  J_1 \cos (q_x) -  J_2 \cos (q_y)
\eea

\begin{figure}[h!]
    \centering
    \includegraphics[scale=0.4]{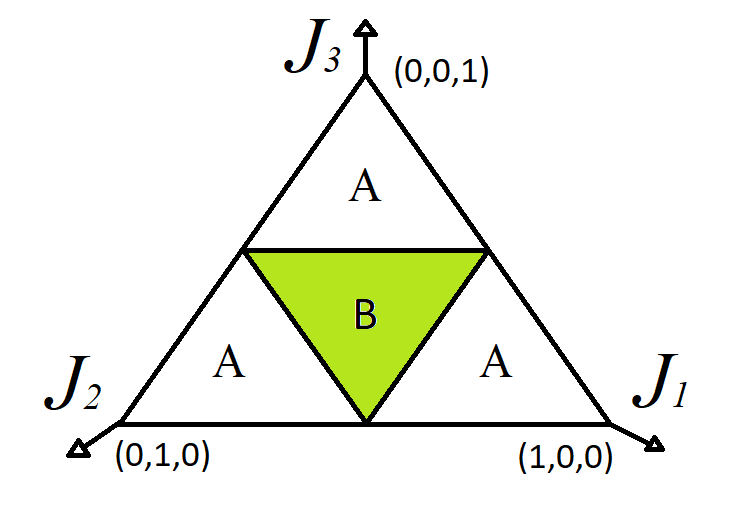}
    \caption{{The phase diagram of Kitaev model in honeycomb lattice. The regions marked A are gapped regions and B is the gapless region.\cite{utso2016dynamical}}}
    \label{kitaev_phased}
\end{figure}

The phase diagram of the system in the vortex free state is given in Fig [\ref{kitaev_phased}]. There is a gapless region for the parameter values satisfying the inequality $|J_1 - J_2|  \le  J_3 \le J_1 + J_2 $ and a gapped region elsewhere. These two phases are topologically different \citep{kitaev2006,feng2007,kitaev2010} and cannot always  be detected by studying Loschmidt echo \cite{Kehrein}.


\begin{figure*}
\includegraphics[scale=.22]{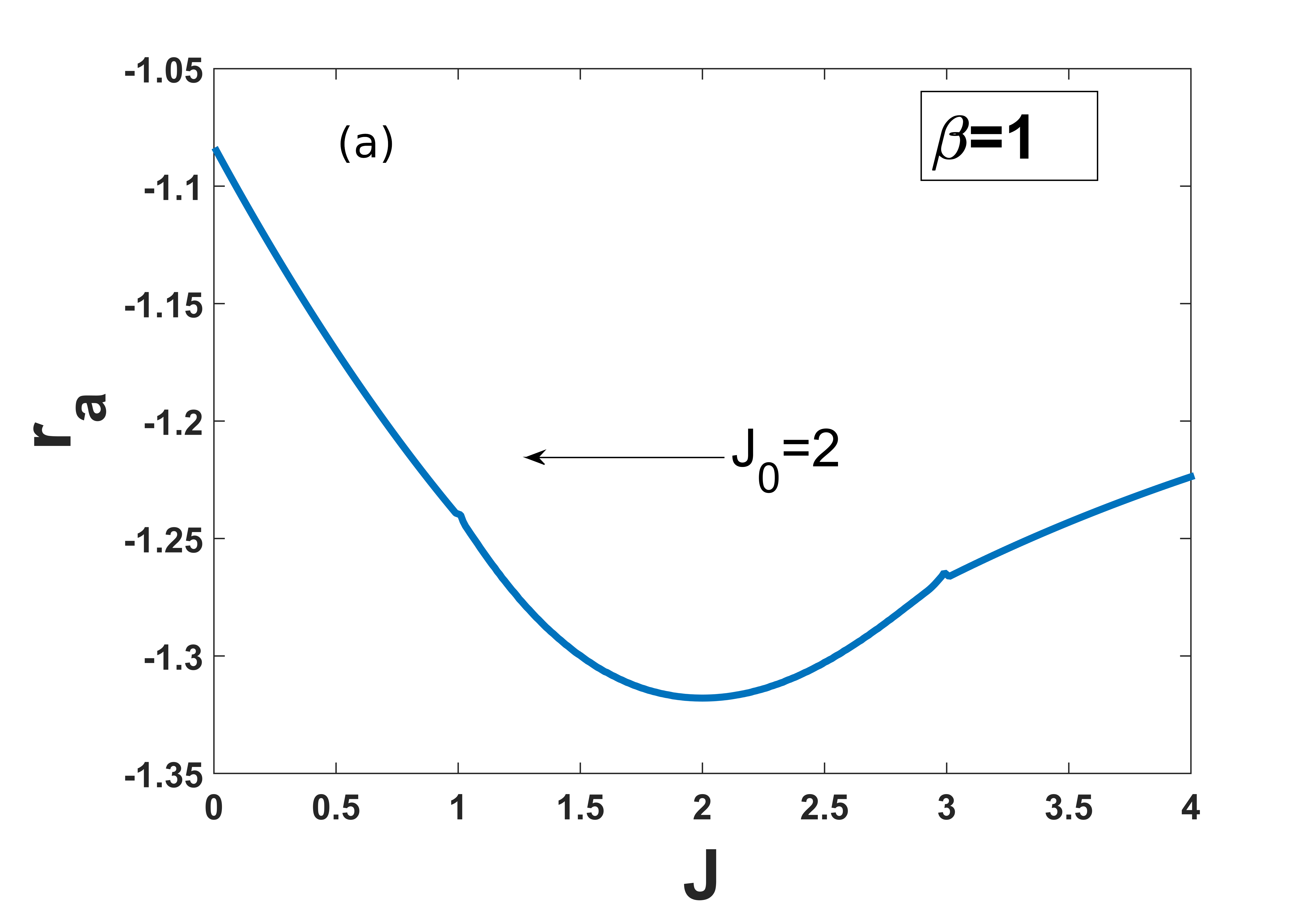}
\includegraphics[scale=.22]{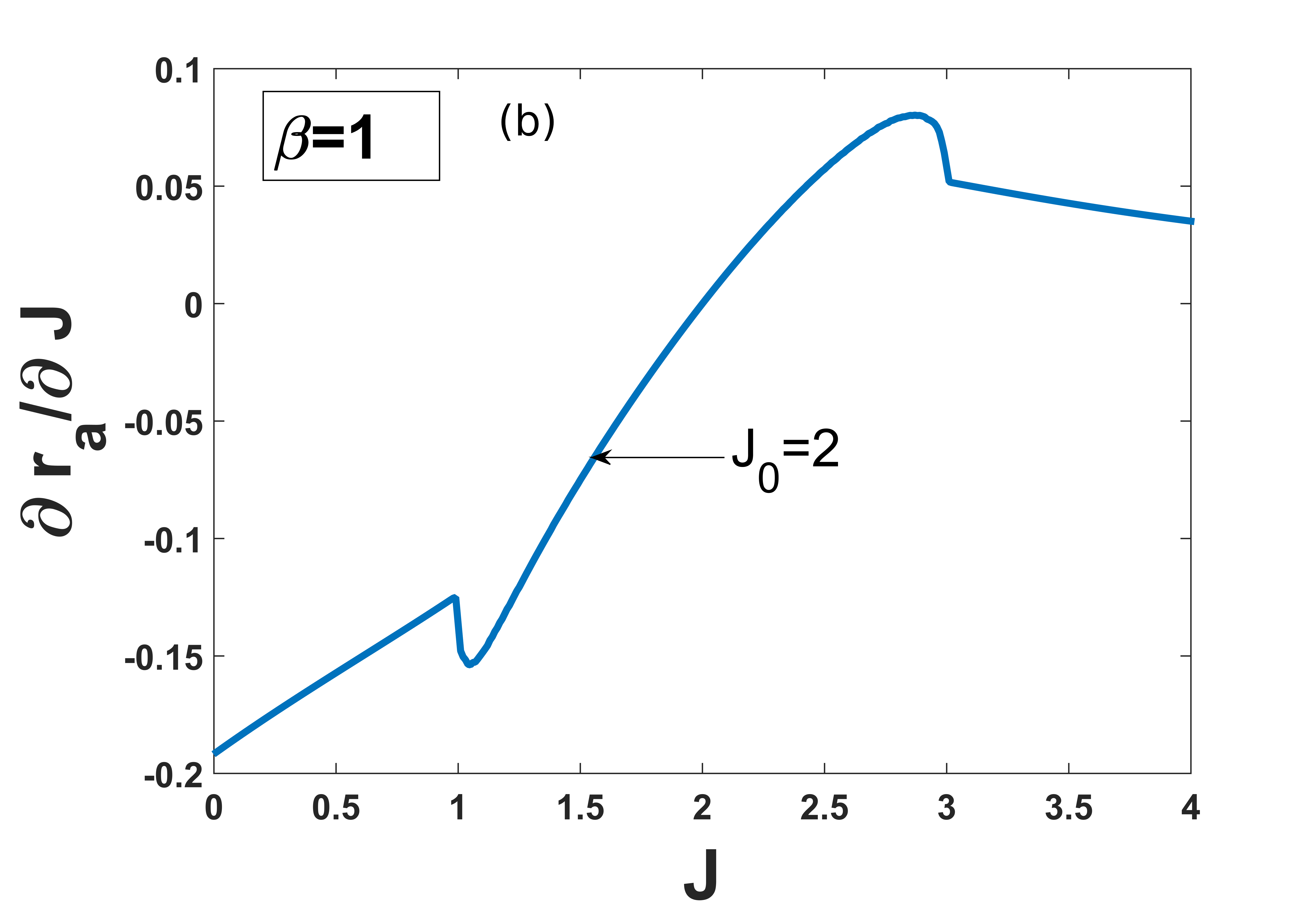}
\includegraphics[scale=.22]{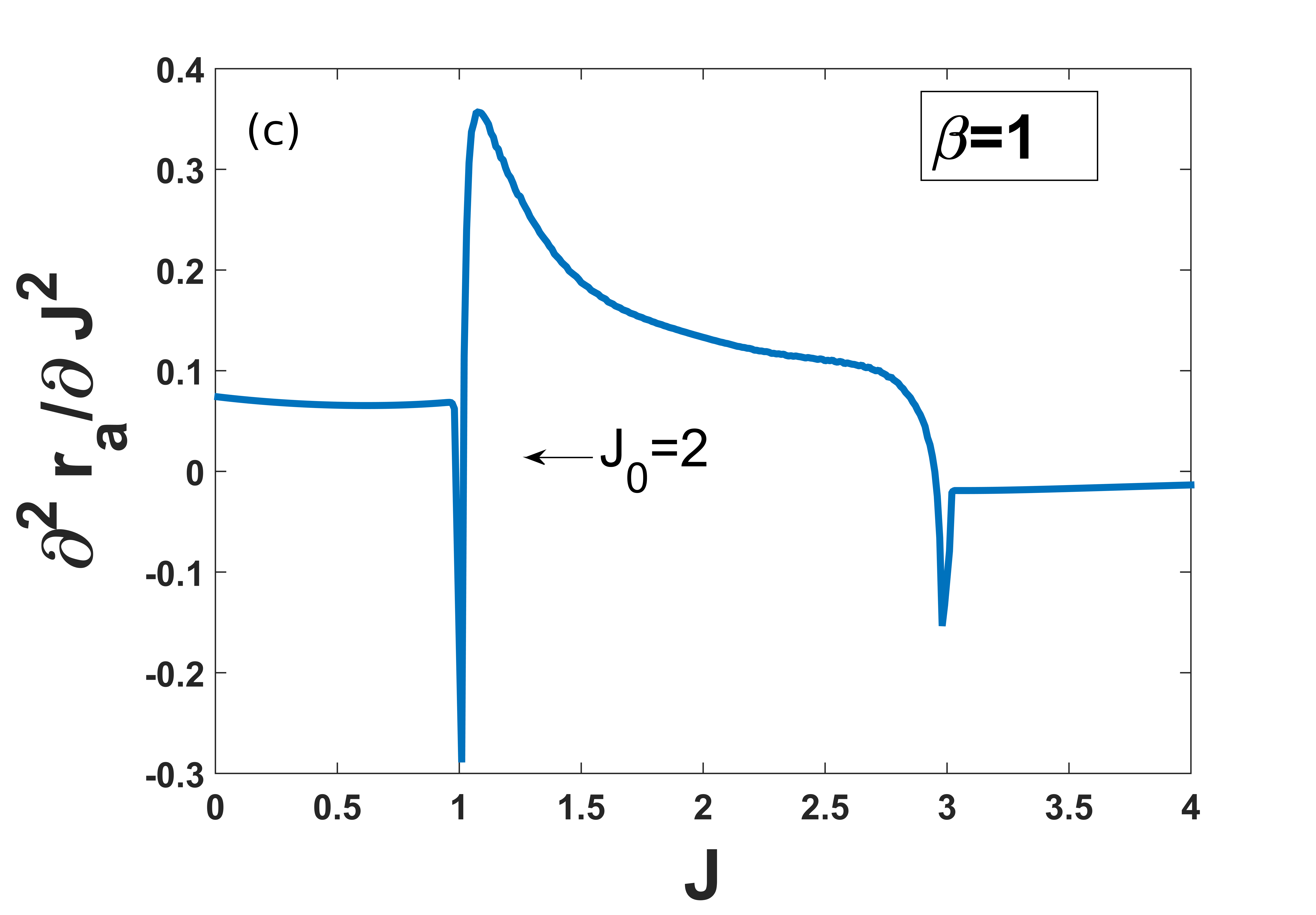}
\includegraphics[scale=.22]{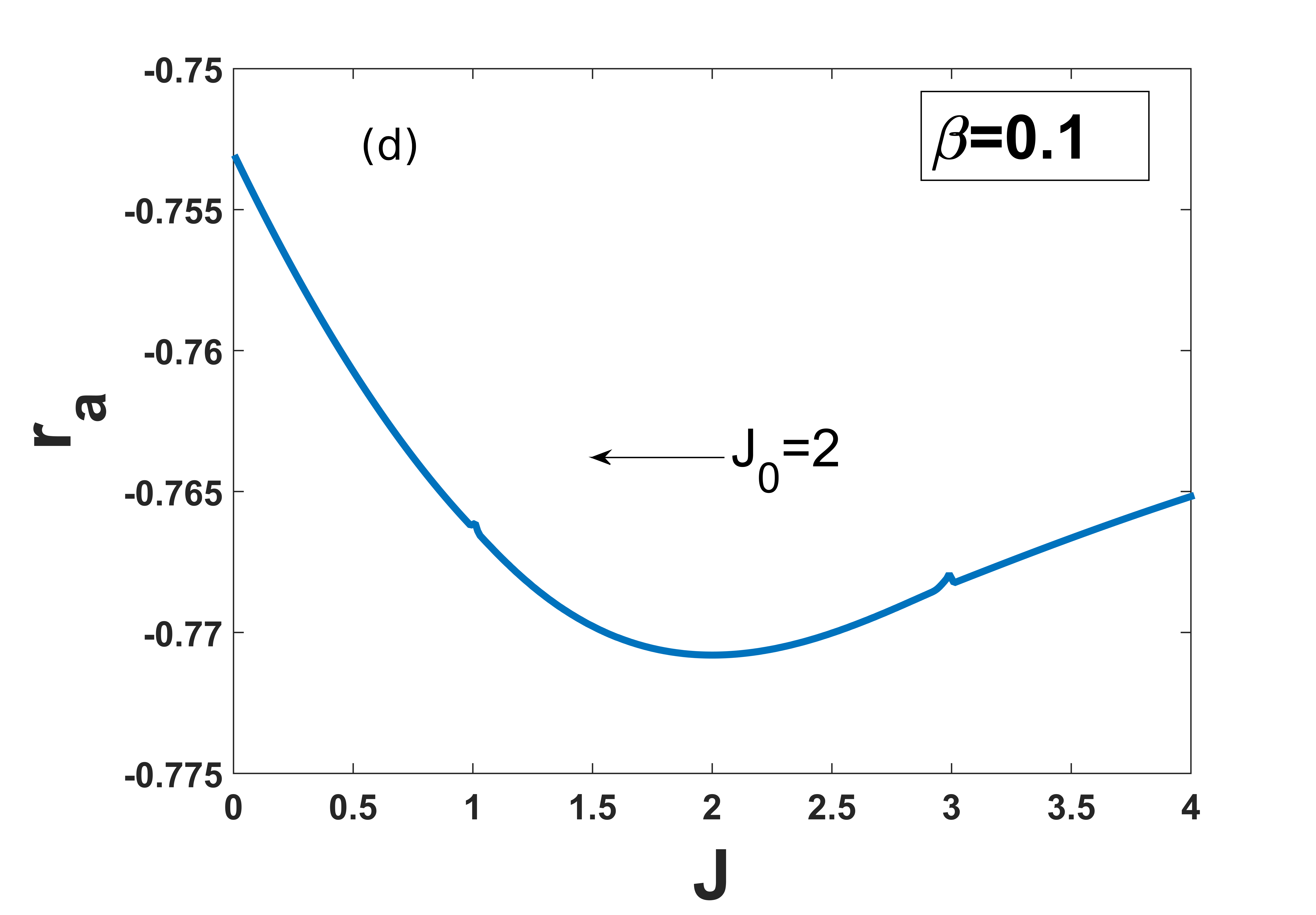}
\includegraphics[scale=.22]{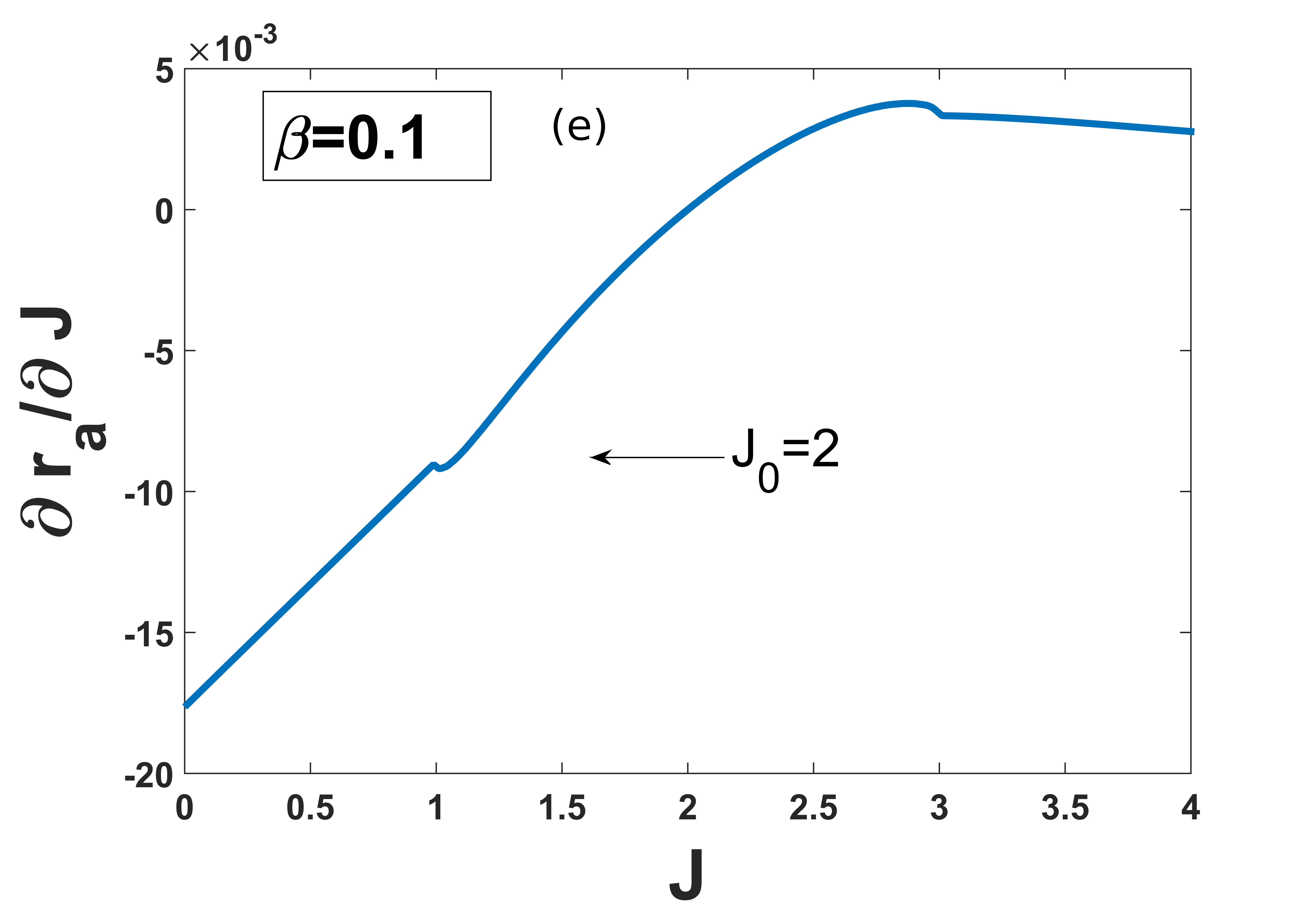}
\includegraphics[scale=.22]{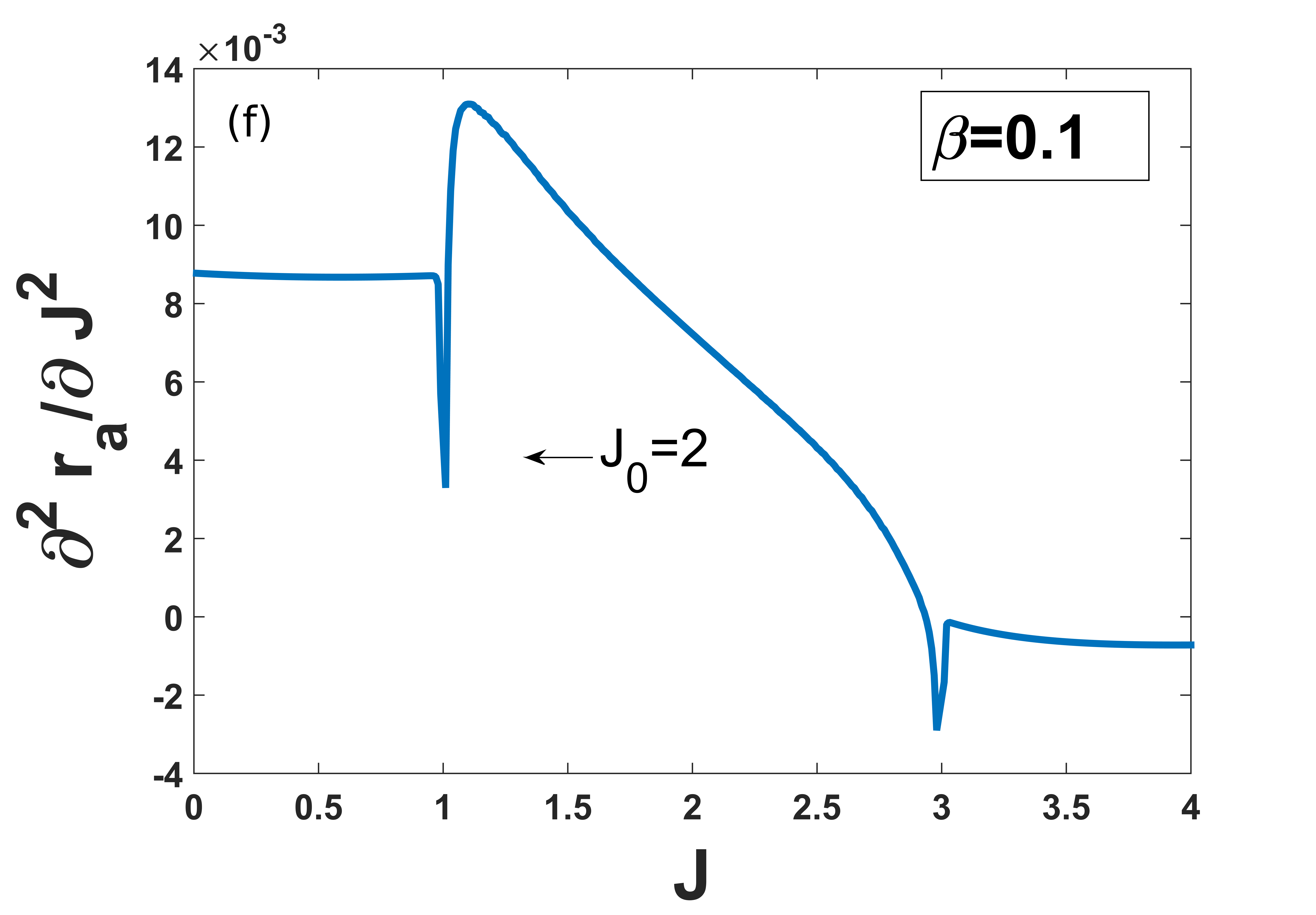}
\includegraphics[scale=.22]{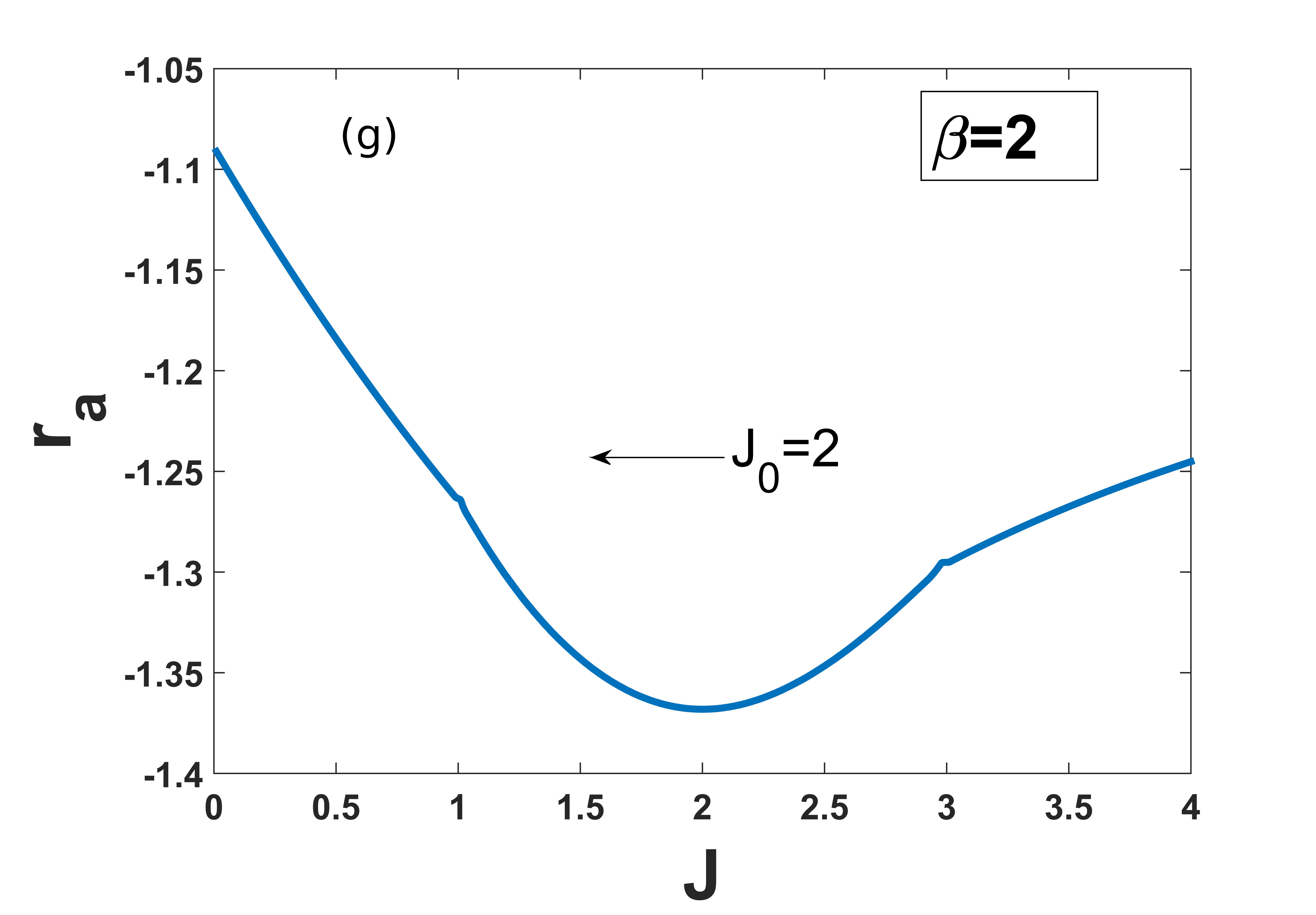}
\includegraphics[scale=.22]{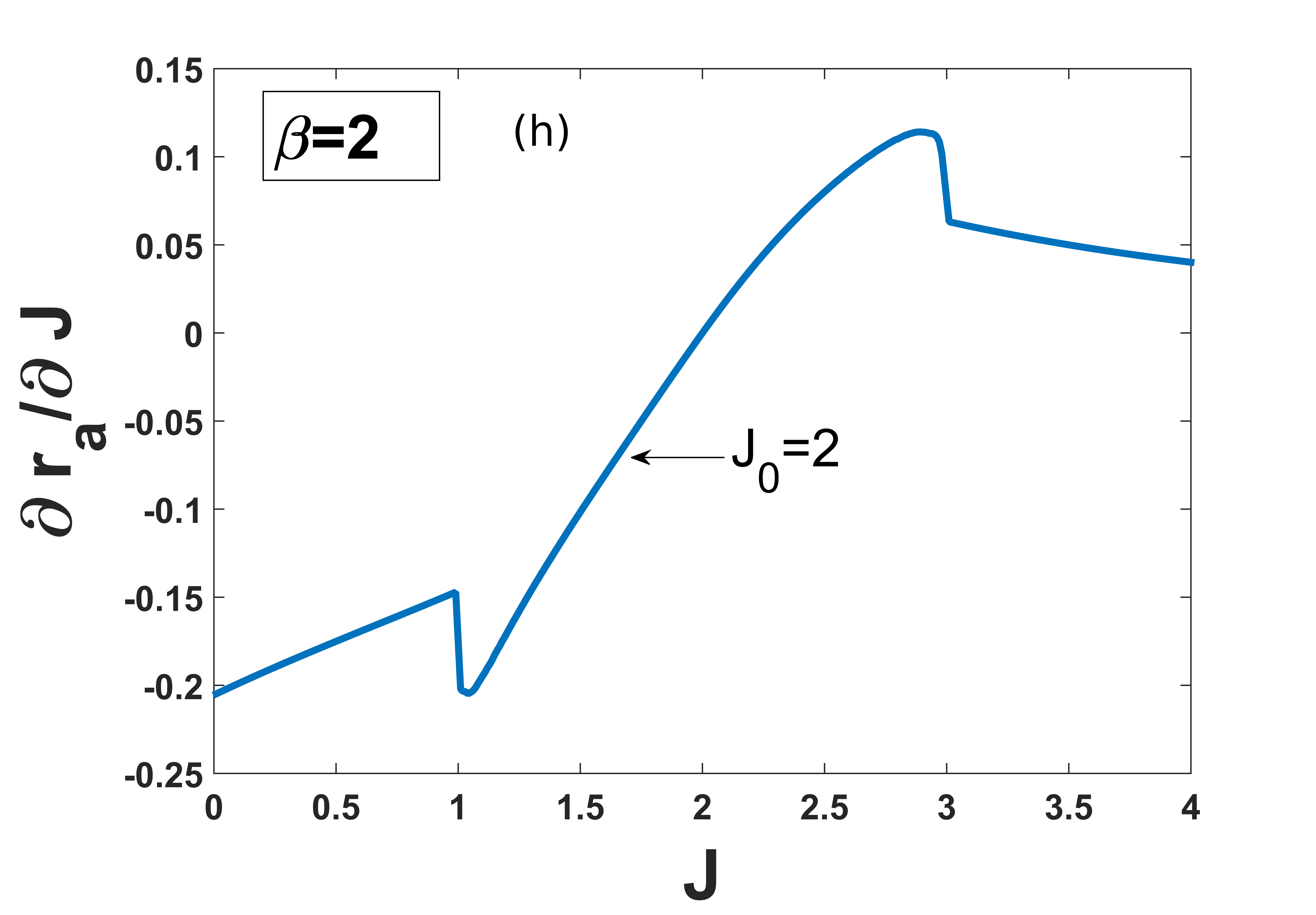}
\includegraphics[scale=.22]{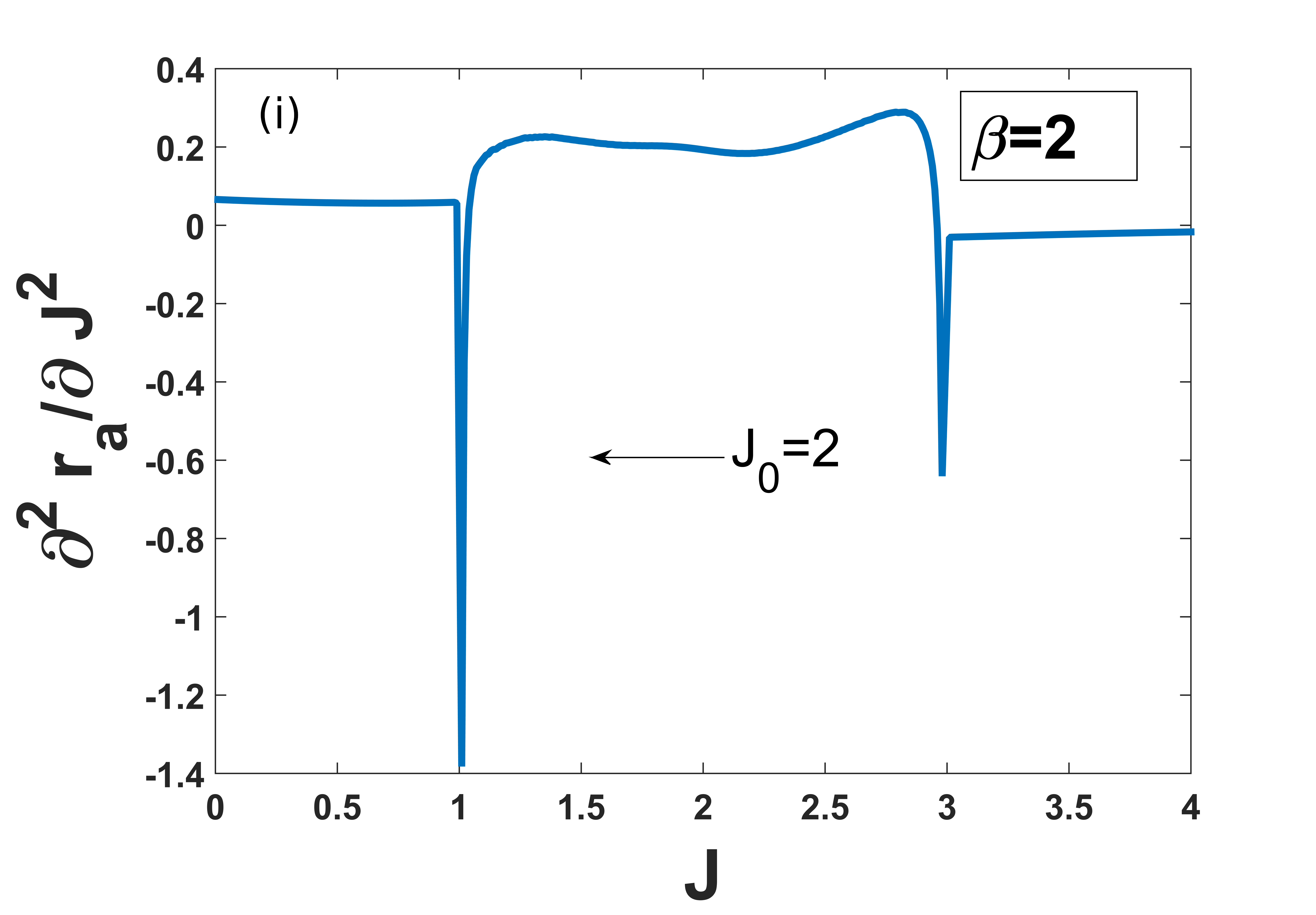}
\caption{Kitaev model on honeycomb lattice: Plots of $r_a$, $\partial r_a/\partial J $, $\partial^2 r_a/\partial J^2$ vs $J$. We have fixed $J_1=2$ and $J_2=1$. We have plotted for three different temperatures $\beta=1$ in (a),(b)and (c), $\beta=0.1$ in (c),(d)and (e) and $\beta=2$ in (g), (h) and (i). The initial value of $J_3$ is $J_0=2$. It is evident from the plots that the rate function is showing nonanalyticities at QCPs at $J=1$ and $J=3$. }
\label{j1nej2_v1}
\end{figure*}

\begin{figure}
\includegraphics[scale=.22]{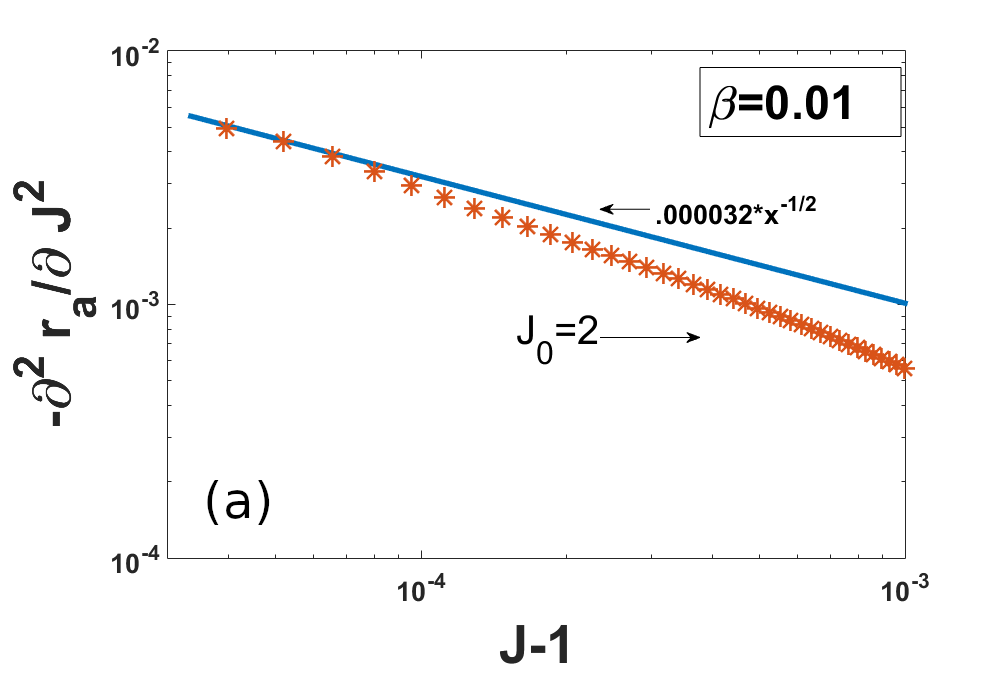}
\includegraphics[scale=.22]{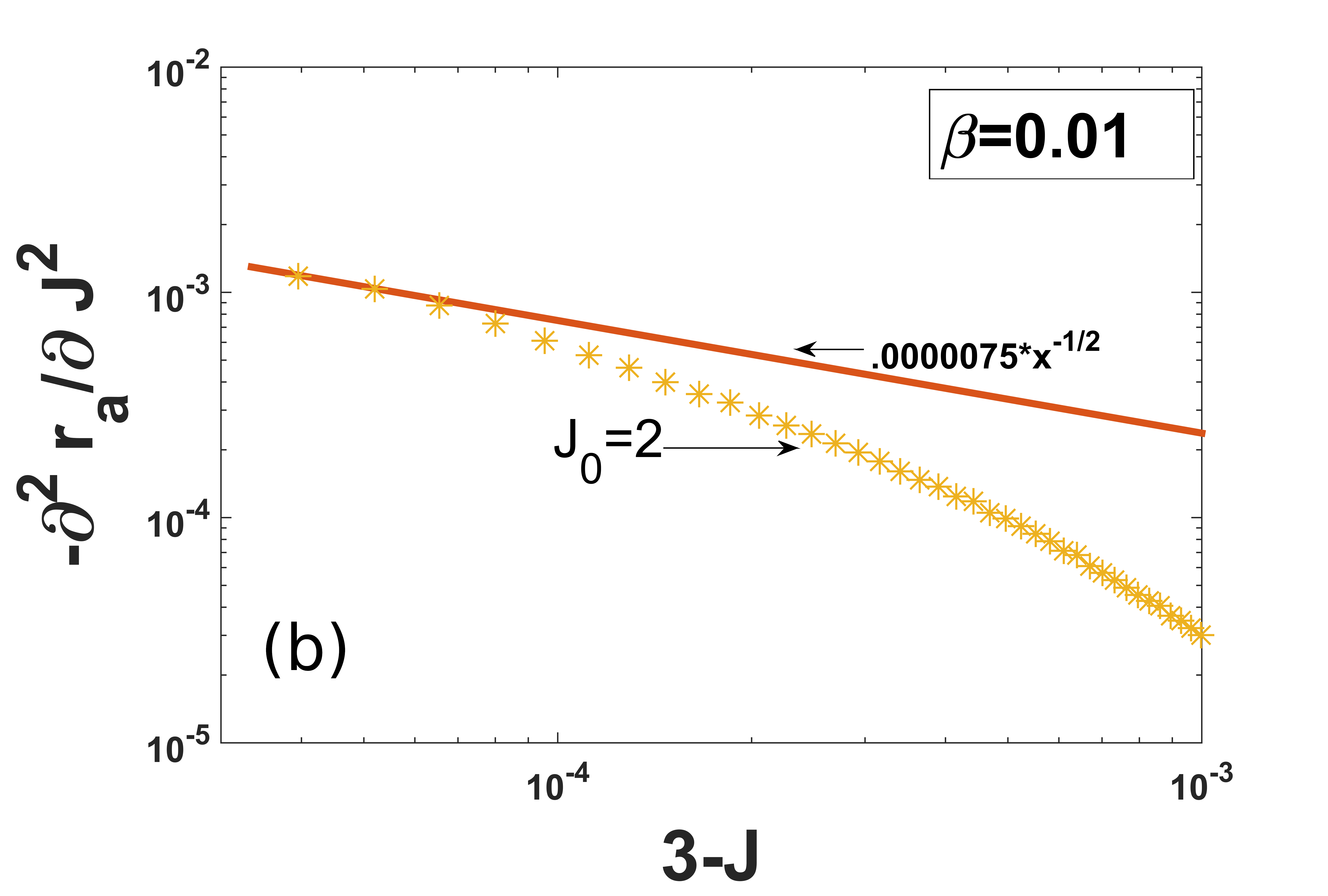}
\caption{Kitaev model on honeycomb lattice: (a) $-\partial^2 r_a/\partial J$ vs $J-1$ is plotted for $J_1=2$ and $J_2=1$. As it approaches critical point $J=1$, it coincides with the blue line which diverges as $x^{-1/2}$. (b) $-\partial^2 r_a/\partial J^2$ vs $3-J$ is plotted. As it approaches critical point $J=3$, it starts diverging as $x^{-1/2}$.}
\label{kit_close}
\end{figure}

We have dealt with the case where $J_1=J_2$ in our previous paper \cite{nandi2022}. Here we will deal with the more general case of $J_1 \ne J_2$. Using Eq. (\ref{rddim}), one can calculate numerically the rate function and its derivatives for quench of the parameter $J_3$ for a fixed $J_1$ and $J_2$ (Fig \ref{j1nej2_v1}). We find that a nonanalyticity occurs at $|J_1-J_2|$ and $J_1+J_2$ and that the double derivative of the rate function diverges with a critical exponent of $1/2$ (Fig. \ref{kit_close}). 

We shall now present an analytic treatment for the case $J_1 \ne J_2$ for a quench of $J_3$ from $J_0$ to $J$.
Here, the parameter $p$ is $J$, $V$ is $4\pi^2$ and $\mathcal{L}_{\v}$ can be written as 
\bea 
\mathcal{L}_{\v}=(J-J_0)^2 \frac{{a'}_{\v}^2}{\lambda_{\v}^2 \; {\lambda'}_{\v}^2} \no
\eea
We substitute 
\[ M=J_1 + J_2, \; N=J_1-J_2,\;u=\frac{q_x+q_y}{2}, \;v=\frac{q_x-q_y}{2}  \]
to get
\bea
 a_{\vec{q}} &=& -M\cos (u) \sin (v) -N\sin (u) \cos (v) \no\\ b_{\vec{q}} &=& J_3-M\cos (u) \cos (v) +N\sin (u) \sin (v)  \label{ab_MNuv}
\eea
 We can now write the the expression of $r_a$ from Eq.(\ref{rlog}). Since only the third term will remain after the differentiation with respect to $J$, we will calculate the integrand in the third term of the rate function, namely,
\bea I_n = 4(J-J_0)^{2n} \int_{u=0}^{\pi/2} du \; \int_{v=-\pi}^{\pi} dv   \;    \left(\frac{a'}{\lambda'}\right)^{2n} \label{In}\eea  

The analytic expression of $I_1$ is obtained as (see Appendix~B)
\bea I_1 &=& {\frac{2\pi}{J}}(J-J_0)^2[2J u_c-(M^2-N^2)\frac{\sin (2u_c)}{2J} \no\\ &+& {\frac{1}{J}}(M^2+N^2)({\frac{\pi}{2}}-u_c)] \eea
where $u_c=\cos^{-1}(\sqrt{(J^2-N^2)/(M^2-N^2)})$.
If we approach $|N|$ from the gapless phase and approximate $J$ close to $|N|$ as $|N|+\epsilon$, we obtain 
\bea \frac{\partial^2 I_1}{\partial J^2} = -\frac{4\pi\sqrt{2}(|N|-J_0)^2}{\sqrt{|N|(M^2-N^2)}}\epsilon^{-1/2} \eea
upto leading order.

In the same way, we can show that if we approach $M$ from the gapless phase and approximate $J$ close to $M$ as $M-\epsilon$,
\bea  \frac{\partial^2 I_1}{\partial J^2} = -\frac{4\pi\sqrt{2}(M-J_0)^2}{\sqrt{M(M^2-N^2)}}\epsilon^{-1/2}  \eea
upto leading order.

We conclude that the double derivative of $I_n$ will also be proportional to $\epsilon^{-1/2}$ upto the leading order. 
Hence the double derivative of $r_a$ diverges algebraically with the exponent $1/2$ near both the critical points at any temperature (Fig \ref{kit_close}). 
 Although we have not considered vortex excitation in the nonzero temperature, the excitation of vortices does not destroy the singular behavior of the rate function $r_a$. The vortex excitation in 2D Kitaev model is adiabatic with temperature and does not induce any phase transition \cite{Nasu}. Hence the vortex-free configuration gives the main contribution to the nonanalytic feature of the rate function at nonzero temperatures.

\section{Three Dimensional Case}\label{3D}
 In the case of applying our detector to 3D Hamiltonians, we consider two types of topological materials, namely, the Weyl Semimetals and the Topological Nodal Line Semimetal. 
\subsection{Weyl Semimetal}    
The commuting Hamiltonians for Weyl semimetals can be written as \cite{Vishwanath2018, Rao2016},

\bea  {\mathcal H}_{\vec{q}}  =  a_{\vec{q}} \sigma_1 +  b_{\vec{q}} \sigma_2 + c_{\vec{q}} \sigma_3 \label{H-WM-def} \eea
where  $a_{\vec{q}} = \sin (q_x)$, $b_{\vec{q}} = \sin (q_y)$, $c_{\vec{q}} =  J_3 - \cos  (q_x) - \cos (q_y) -\cos (q_z)$ and $\vec{q}$ runs over a simple cubic lattice in the range $(-\pi<q_x,q_y,q_z<\pi)$. 

The ground state of this Hamiltonian shows a gapless phase for $J_3<3$ and a gapped phase for $J_3>3$. We consider a quench $J_3=J_0 \to J_3=J$. 
\begin{figure}
\includegraphics[scale=0.33]{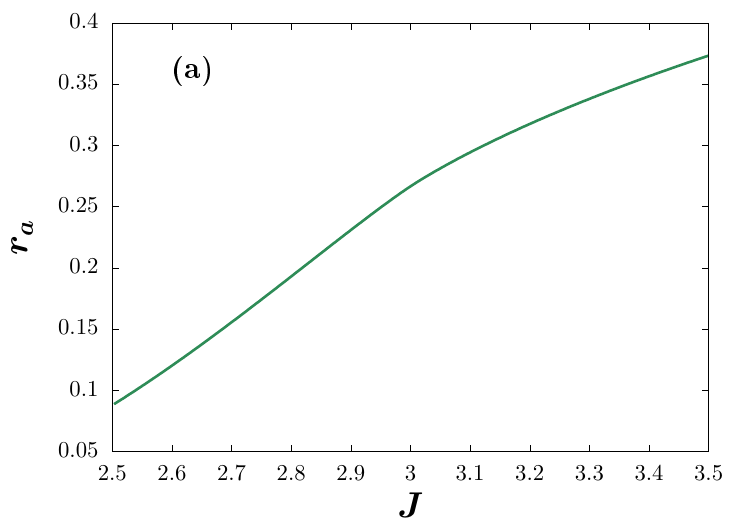}
\includegraphics[scale=0.33]{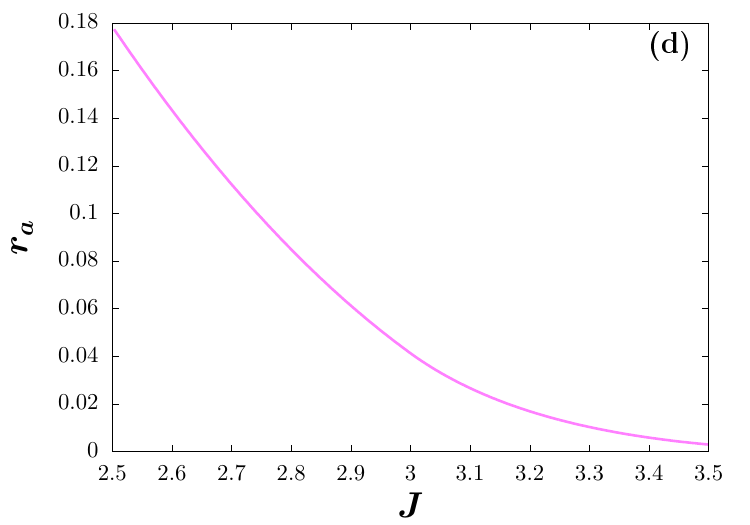}
\includegraphics[scale=0.33]{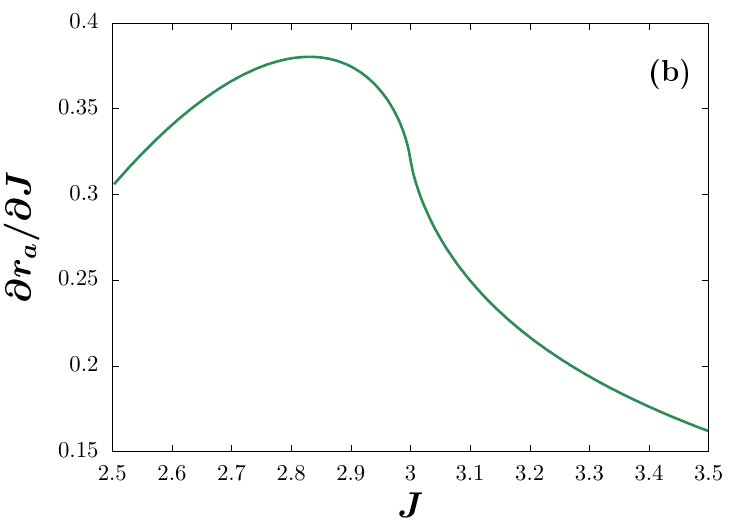}
\includegraphics[scale=0.33]{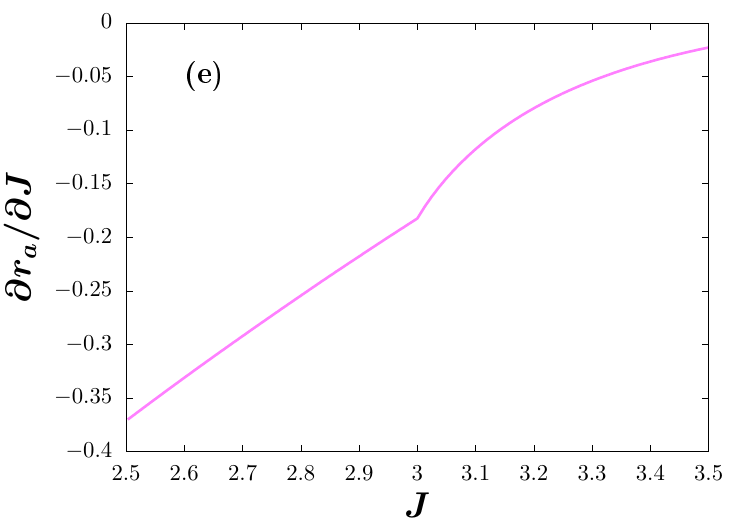}
\includegraphics[scale=0.33]{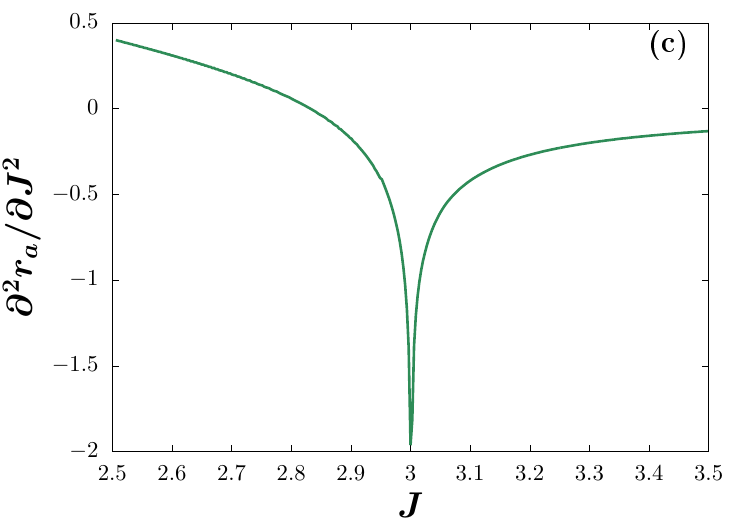}
\includegraphics[scale=0.33]{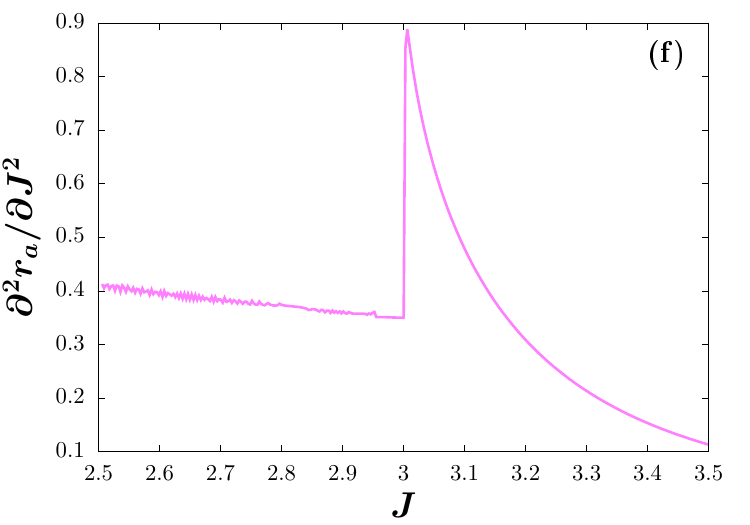}
\caption{ 3D Weyl semimetal: Rate function $r_a$ and its two derivatives computed numerically from Eqs. (\ref{rddim}) for quench of the parameter $J_0 \to J$  at $\beta=4$. For (a), (b) and (c), the pre-quench value is within the gapless phase $J_0=2$, while for (d), (e) and (f) it is within the gapped phase $J_0=3.8$. In both cases the quantity $\partial^2 r_a/\partial J^2$ diverges at the phase boundary $J=3$. }
\label{Weyl_Fig}
\end{figure}

\begin{figure}
\includegraphics[scale=.22]{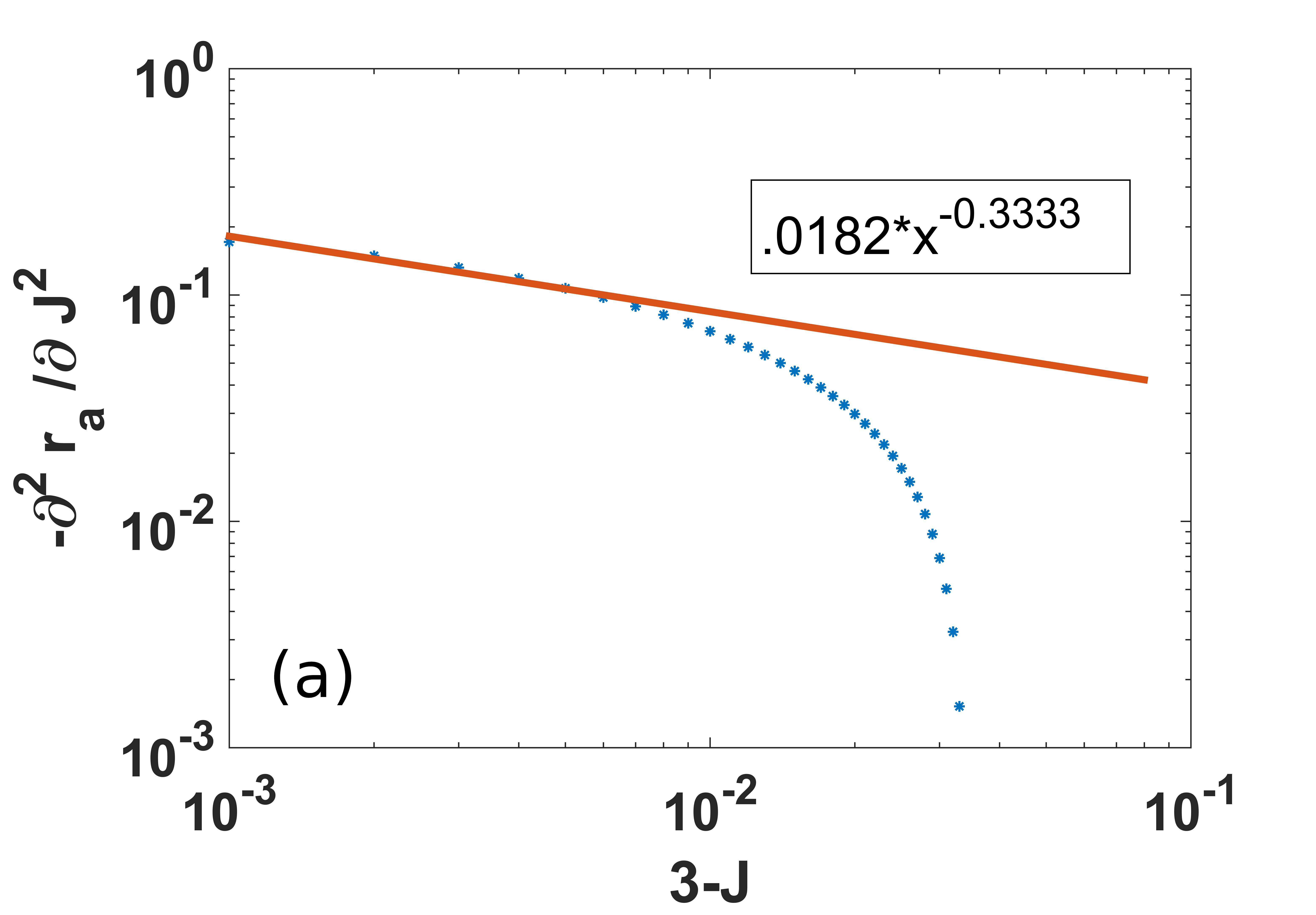}
\includegraphics[scale=.22]{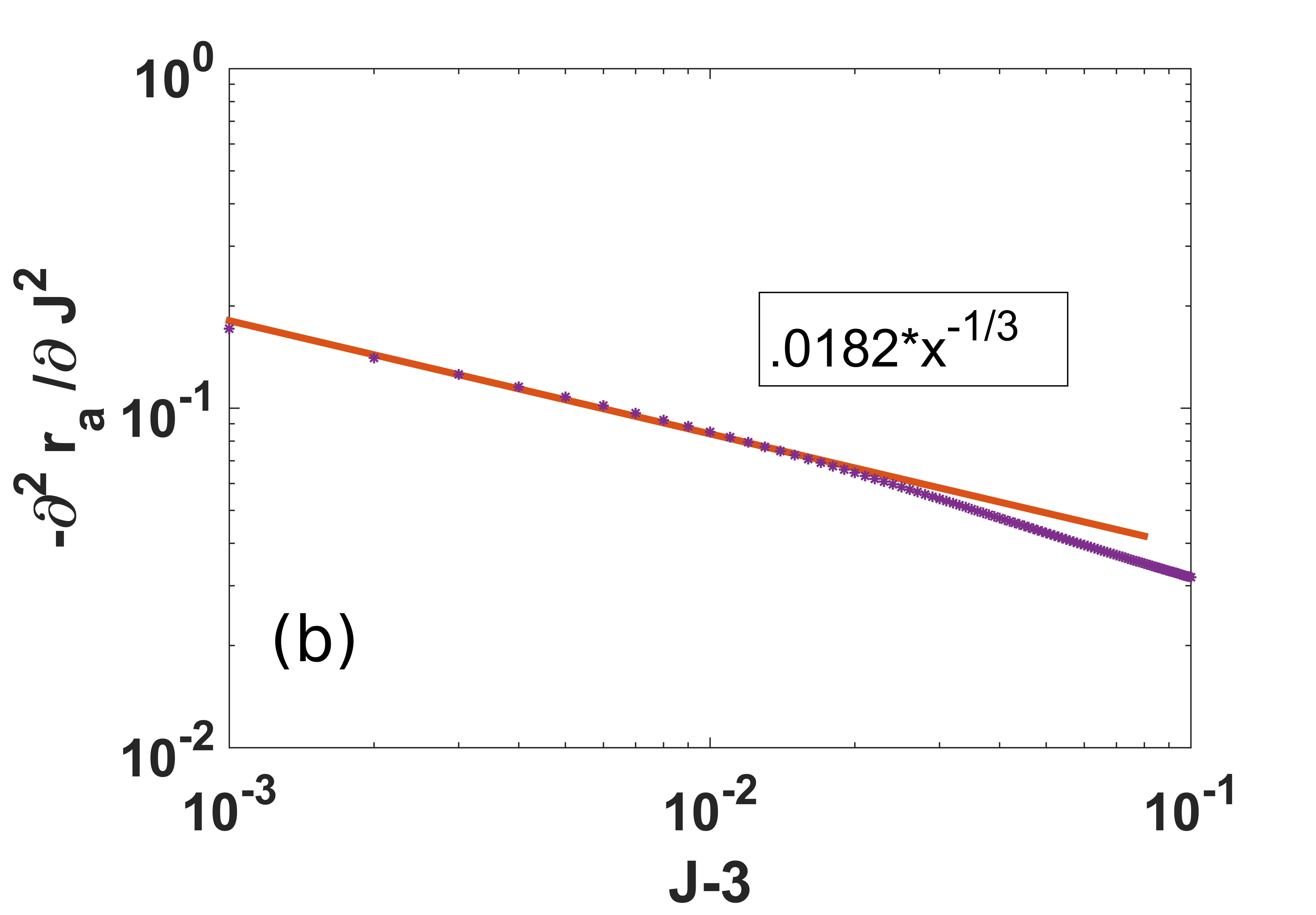}
\caption{3D Weyl Semimetal:(a) $-\partial^2 r_a/\partial J$ vs $J$ is plotted very close to the quantum critical point at $J=3$. It is evident from the figure that double derivative diverges algebraically both from left(a) and from right(b) of the critical point with a critical exponent $-1/3$.}
\label{weyl_close}
\end{figure}

 We numerically evaluate the rate  function  $r_a$ from Eq.~(\ref{rddim}) 
\bea r_a(\beta,J_0,J) &=& 3\log 2-\frac{1}{8\pi^3}\int_{\vec{q}} \log(1+\alpha_{\vec{q}}) d\vec{q}\nonumber \\ &-&\frac{1}{4\pi^3} \int_{\vec{q}} \log \left[1+\sqrt{1-\gamma_{\vec{q}} \mathcal{L}_{\vec{q}}}\right] d\vec{q} \label{Weyl1} \eea
where $p=J$, $V = 8\pi^3$ and $\mathcal{L}_{\vec{q}}$ for Weyl Semimetal is 
\be \mathcal{L}_{\vec{q}}=(J-J_0)^2 (\sin^2 (q_x)+\sin^2 (q_y))/(\lambda_{\vec{q}}\lambda'_{\vec{q}})^2\ee
We observe that the first derivative (with respect to $J$) of $r_a$ shows a change of slope at the QCP $J=3$ both for $J_0 < 3$ and $>3$ (Fig. \ref{Weyl_Fig}) and the double derivative diverges algebraically with a critical exponent $1/3$ (Fig. \ref{weyl_close}). It is important to mention that, unlike the previous two cases, this singularity is visible only at low temperatures and we could not study the behaviour of $r_a$ analytically.

\subsection{Topological Nodal Line Semimetal} When  $b_{\vec{q}} =0$ in Eq. (\ref{H-WM-def}), one has a topological nodal line semimetal \cite{Okugawa}. We write the commuting Hamiltonians as 
\bea \mathcal{H}_{\vec{q}} = a_{\vec{q}}\sigma_1 + c_{\vec{q}}\sigma_3 \label{our_tnlsm_def}\eea
where $c_{\vec{q}}=J_3-\cos (q_x) -\cos (q_y) -\cos (q_z)$ and $a_{\vec{q}}=\sin (q_z)$.
Usually one uses the Hamiltonian 
\be 
{\mathcal H}_{\vec{k}}^0 = v k_z \sigma_x + (k^2 - k_0^2) \sigma_z \label{usual_Hk} 
\ee
(with $v$ and $k_0$ as parameters) for this type of materials \cite{Burkov2011, Armitage2018, Yang2022}. The Hamiltonian of Eq. (\ref{our_tnlsm_def}) reduces to this form on the plane $k_x = -k_y$, upto terms quadratic in $\vec{k}$.

The Hamiltonian (\ref{our_tnlsm_def}) has a gapped phase in the region $J_3>3$ while it shows a gapless phase  when $J_3<3$. For a quench $J_3=J_0$ to $J_3=J$, the rate function $r_a$ is calculated from Eq (\ref{Weyl1}) with   
\bea \mathcal{L}_{\vec{q}}=(J-J_0)^2(\sin (q_z))^2/(\lambda_{\vec{q}}\lambda'_{\vec{q}})^2 \eea

\begin{figure}
\includegraphics[scale=0.33]{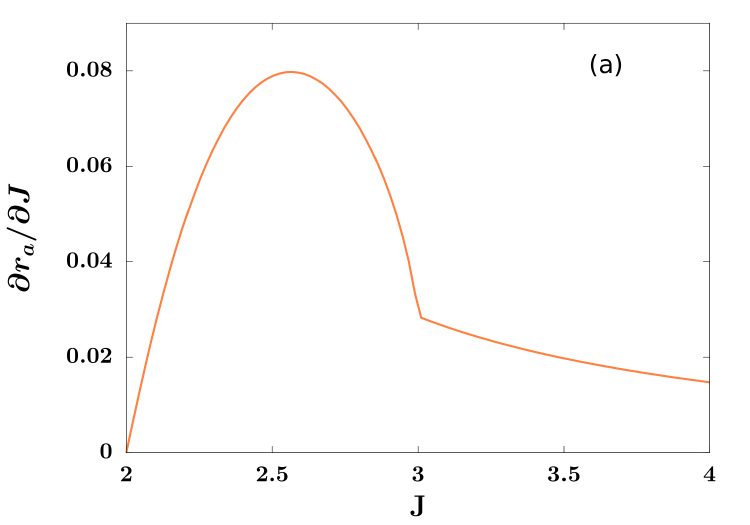}
\includegraphics[scale=0.33]{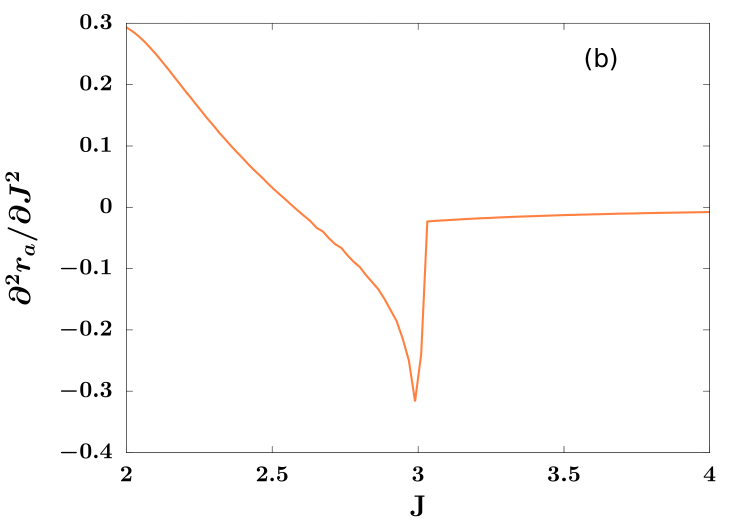}
\includegraphics[scale=0.33]{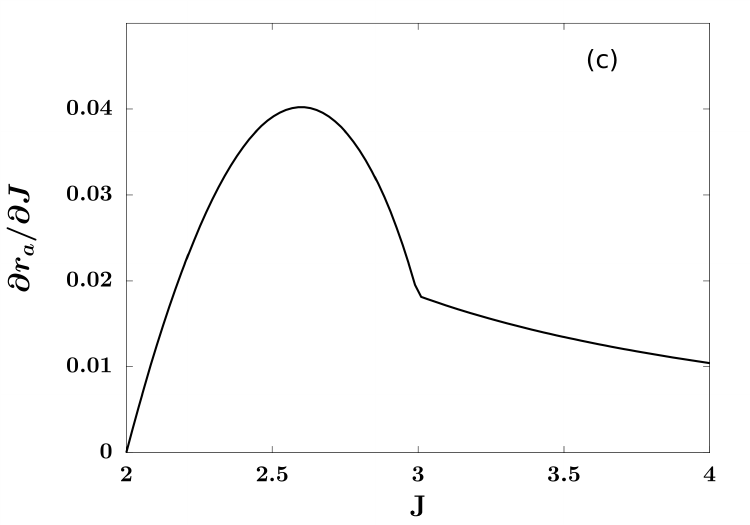}
\includegraphics[scale=0.33]{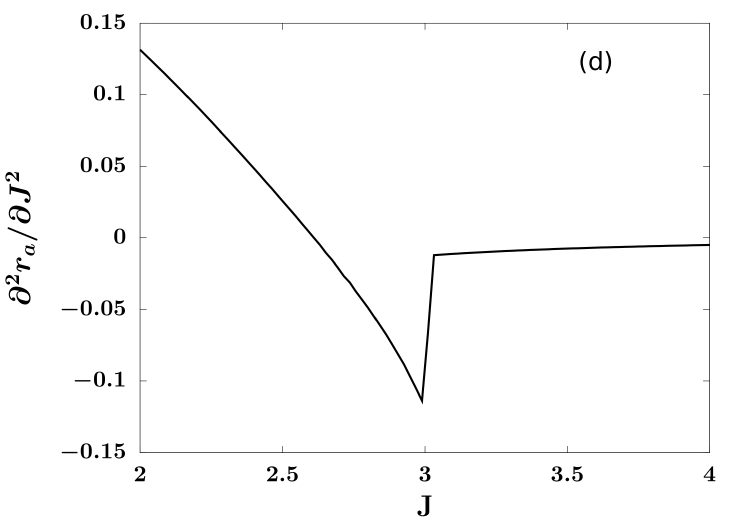}
\caption{3D topological nodal line semimetal: First and second derivatives of the rate function computed numerically from Eqs. (\ref{rddim}) for quench of the parameter $J_0=2 \to J$  at $\beta=4$ [(a) and (b)] and $\beta=1$ [(c) and (d)]. In both cases the quantity $\partial^2 r_a/\partial J^2$ is discontinuous at the phase boundary $J=3$. }
\label{Toy3d_fig}
\end{figure}

\begin{figure}
    \centering
    \includegraphics[scale=0.35]{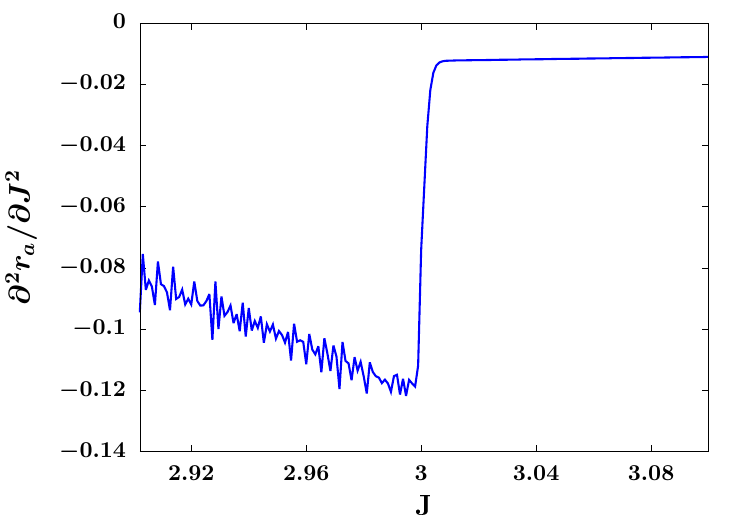}
    \caption{3D topological nodal line semimetal: Double derivative w.r.t J vs J is plotted very close to the quantum critical point at J = 3 for $\beta = 1$ and $J_0 = 2$. It is evident from the figure that there is a finite discontinuity. The reason behind the oscillation in the curve is computational limitation.}
    \label{tnlsmclose}
\end{figure}

In this case also, one observes a change of slope at $J=3$ of the curve $\partial r_a/\partial J$ vs $J$.(Fig. {\ref{Toy3d_fig}}) The second derivative does not show any divergence at the critical point but has a finite discontinuity (Fig. \ref{tnlsmclose}). It is important to mention that, the discontinuity becomes smaller as temperature increases and we could not study the behaviour of $r_a$ analytically in this case.

\subsection{A Topological Insulator Bi$_4$Br$_4$} 
Single layer Bi$_4$Br$_4$ is a quantum spin Hall insulating material. It has been shown that topological edge states persist in multilayer Bi$_4$Br$_4$ even at room temperature
\cite{Yao2015, Hasan2022}
and that it is a suitable candidate for a room temperature topological insulator for its large band gap \cite{Yao2015,Zhou2014,Liu2016,Hsu2019,Li2019,Yoon2020}. 
A low energy effective $k \cdot p$ Hamiltonian for a single layer has been suggested for this material \cite{Yao2015}, and in this section we shall show that our rate function computed using this Hamiltonian correctly predicts the phase transition. \\

\begin{figure}
\includegraphics[scale=0.4]{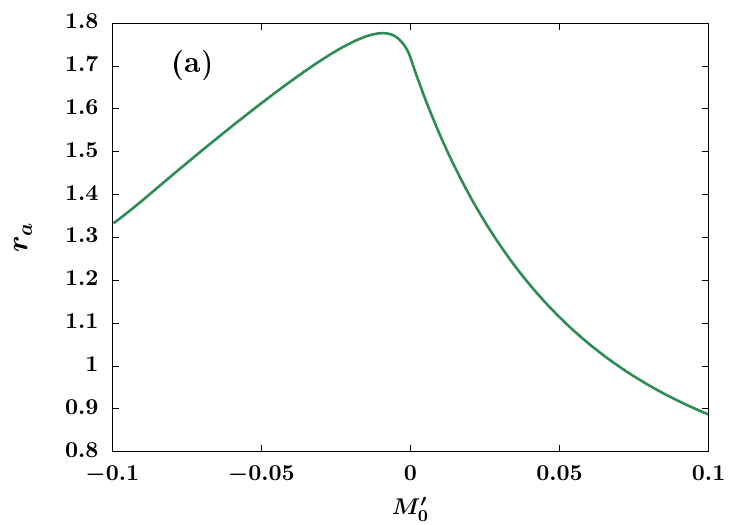}
\includegraphics[scale=0.4]{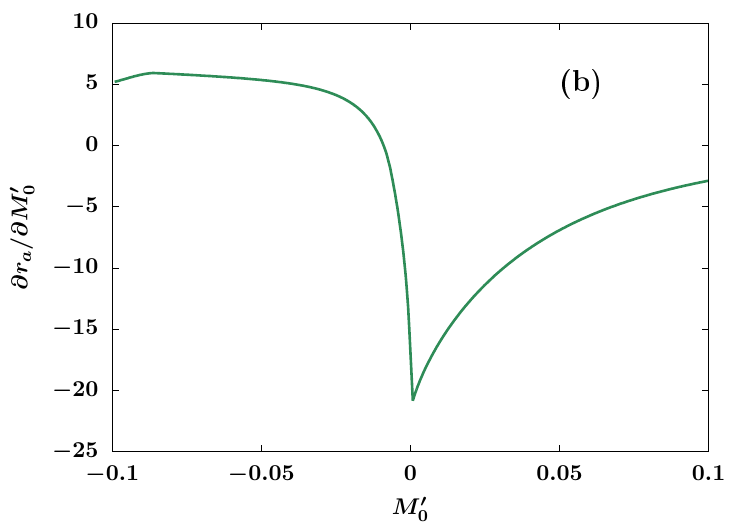}
\includegraphics[scale=0.4]{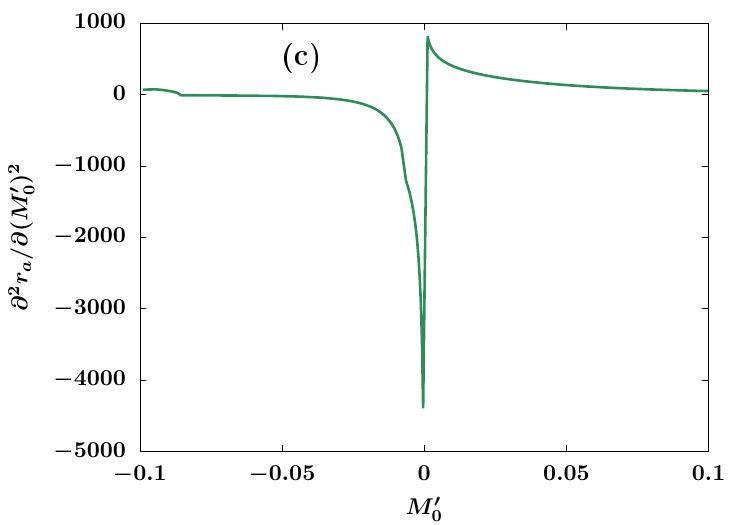}
\caption{Rate function $r_a$ (a) and its first (b) and second (c) derivatives are plotted against $M'_0$ for room temperature. 
Since we have used a $k.p$ Hamiltonian, the integration is done on a region close to the centre of the Brillouin zone, namely, $0.05 \mathring{A}^{-1}\leq q_x,q_y \leq 0.05 \mathring{A}^{-1}$ following \cite{Yao2015}.}
\label{Fig_Bi4Br4}
\end{figure}

The suggested Hamiltonian is 
\begin{eqnarray}
\mathcal{H} &=&
\begin{bmatrix}
M & A_1 q_x & 0 & A_2 q_y \\
A_1^* q_x & -M & A_2 q_y & 0 \\
0 & A_2^* q_y & M & -A_1^* q_x \\
A_2^* q_y & 0 & -A_1 q_x & -M
\end{bmatrix} \nonumber \label{bmatrix}\\
\end{eqnarray}
where 
\begin{eqnarray}
M=M_0 - B_1 q_x^2 - B_2 q_y^2   
\end{eqnarray}
and $A_1=-1.81 + i0.0461$ $eV.\mathring{A}$, $A_2=-4.15 + i0.141$ $eV.\mathring{A}$, $B_1=3.86$ $eV.\mathring{A}^2$, $B_2=0.0032 $ $eV.\mathring{A}^2$. We will quench the parameter $M_0$ from $M_0$ to $M'_0$ as the quantum critical point is at $M_0=0 eV$. Using the fact that the square of this Hamiltonian is a scalar times unit matrix, one can calculate the exponentials for pre-quench and post-quench Hamiltonian, and 
carry on the calculation of rate function following Sec.~\ref{Theory}.
The result is an equation similar to  Eq (\ref{Weyl1}):
\bea r_a(\beta,J_0,J) &=& 4\log 2-\frac{1}{V}\int_{\vec{q}} \log(1+\alpha_{\vec{q}}) d\vec{q}\nonumber \\ &-&\frac{2}{V} \int_{\vec{q}} \log \left[1+\sqrt{1-\gamma_{\vec{q}} \mathcal{L}_{\vec{q}}}\right] d\vec{q} 
\label{bi4br4eq}
\eea
with
\begin{eqnarray}
\mathcal{L}_{\vec{q}}& =& \frac{1}{2}-\frac{X}{2\lambda^2 \lambda'^2} \nonumber \\
X& =& (MM'+|A_1|^2 q_x^2 + |A_2|^2 q_y^2)^2 - |A_1|^2 q_x^2(M'-M)^2 \nonumber \\ 
&-& |A_2|^2 q_y^2 (M'-M)^2) \nonumber \\
M'&=&M'_0-B_1 q_x^2 -B_2 q_y^2 \nonumber 
\end{eqnarray}
where $V$ is the area of the region of integration close to origin (see Fig. (\ref{Fig_Bi4Br4})).
The numerically computed rate function and its derivative as obtained from Eq (\ref{bi4br4eq}) are plotted against the post-quench value $M'_0$ in Fig. (\ref{Fig_Bi4Br4}). It shows a divergence in its second derivative at $M'_0=0$ at room temperature. Divergence in second derivative can also be obtained at other temperatures. 
This shows that our procedure works for the material Bi$_4$Br$_4$, in spite of the fact that the Hamiltonian is made up of $ 4\times 4$ commuting matrices instead of $2 \times 2$ ones (see Eqs. \ref{genf}, \ref{bmatrix}). 

\section{Zero temperature behaviour} \label{t0case}
At zero temperature, the rate function in Eq (\ref{r}) reduces to the rate function used to define dynamical quantum phase transition \cite{Heylprl} using Loschmidt echo. The system undergoing dynamical phase transition shows nonanalytic peaks when rate function is plotted against time. The seminal idea was that a system undergoes dynamical quantum phase transition only if the system is quenched across the equilibrium quantum critical point. Transverse Ising Model shows such behaviour. But later it has been shown that there may not be nonanalyticity in the rate function versus time plot when the quench is across the critical point or there may be a nonanalyticity when the quench is not across the critical point.\cite{Vajnaprb} \cite{Kehrein} \cite{PhysRevB.99.174311}.
We show here that the long time limit of our rate function at $T=0$ shows nonanalyticity only if the system is quenched across the critical point. Our observation holds for all the Hamiltonians considered in this paper. 
The long time limit of the rate function (\ref{rddim}) at zero temperature will become 
\bea r_a= 2\log 2 - \frac{2}{V}\int_{\vec{q}} \log [1+\sqrt{1-\mathcal{L}_{\vec{q}}}] d\vec{q} \label{grate_T0}\eea

\begin{figure}
\includegraphics[scale=0.33]{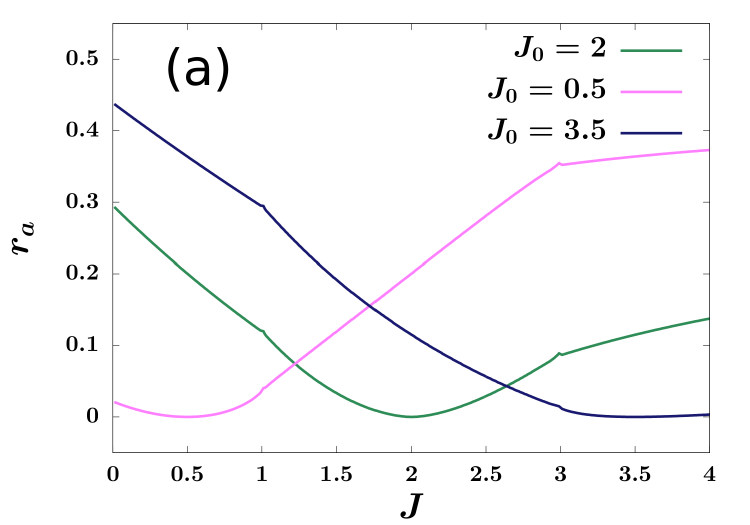}
\includegraphics[scale=0.33]{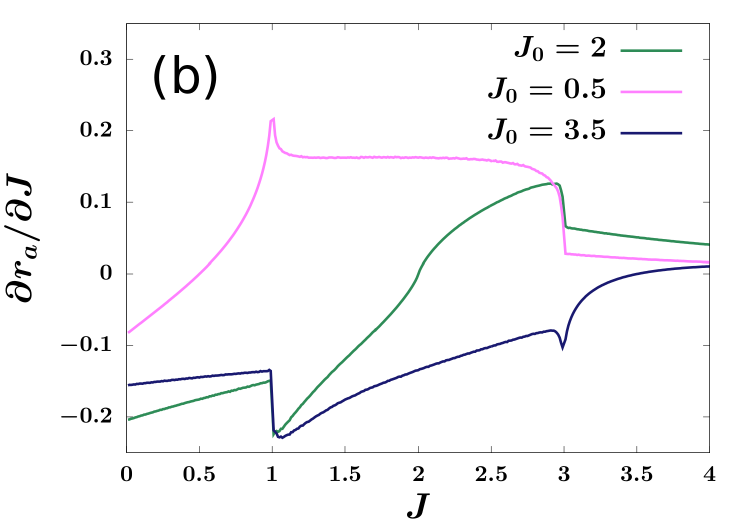}
\caption{ Rate function $r_a$ (a) and its derivative (b) for 2D Kitaev Model computed from Eq. (\ref{grate_T0}) at zero temperature with the pre-quench parameter $J_0=2$ (in gapless phase),$J_0=0.5$ and $J_0=3.5$  (in gapped phase). Nonanalyticity appears only at the phase boundary $J=3$ and $J=1$.}
\label{T0_Results1}
\end{figure}

\begin{figure}
\includegraphics[scale=.16]{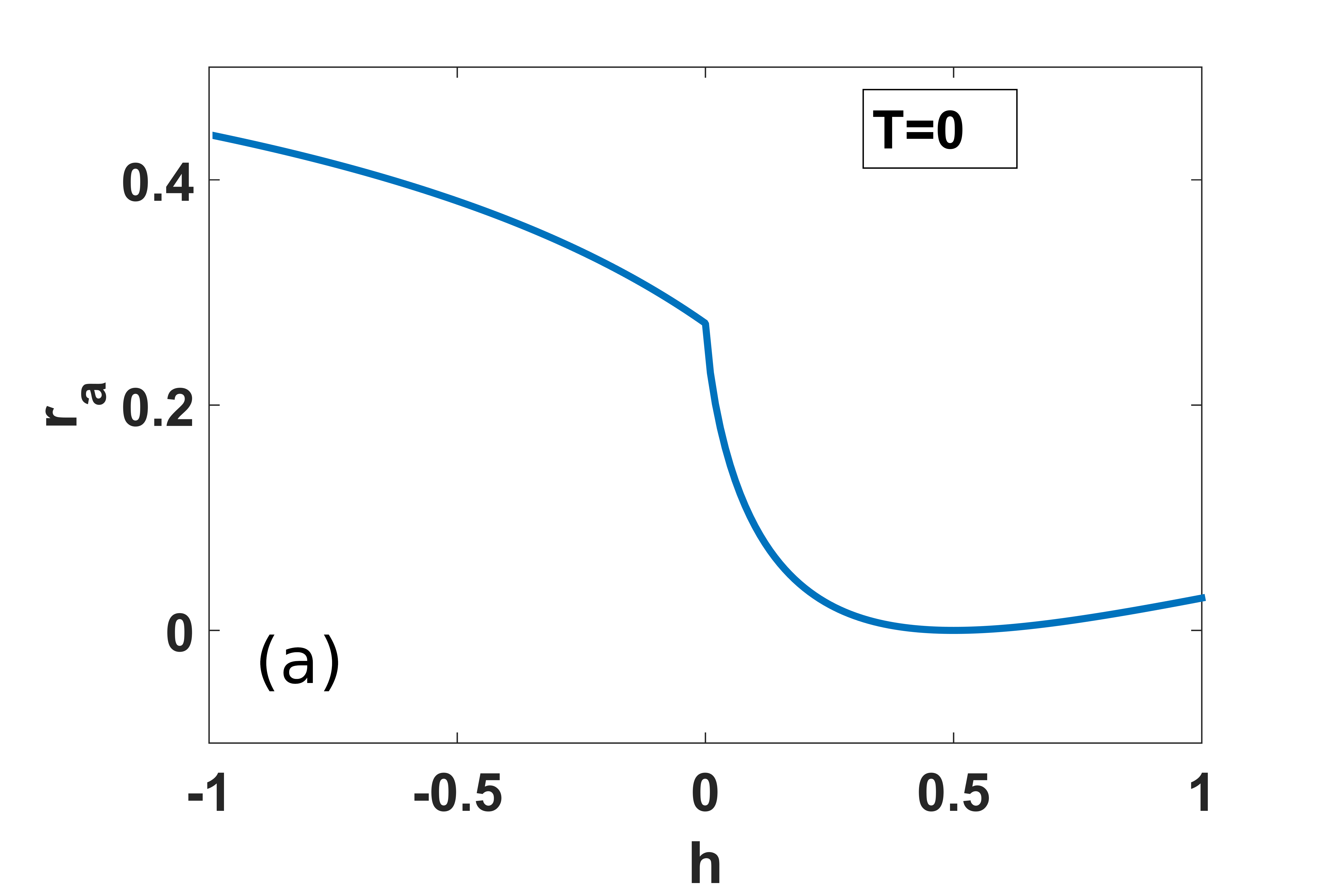}
\includegraphics[scale=.16]{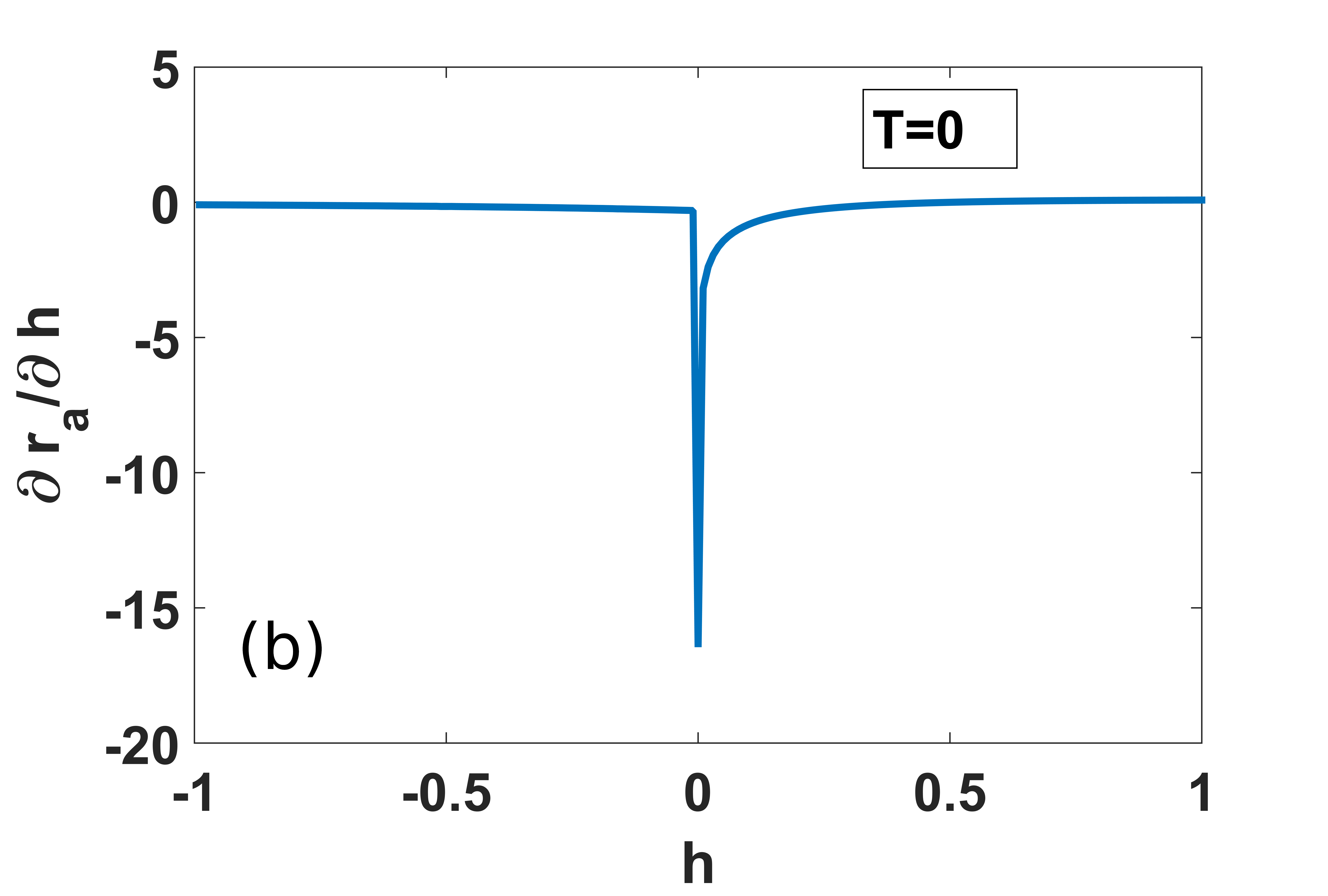}
\caption{Rate function $r_a$ (a) and its derivative (b) for 1D XY Model computed from Eq. (\ref{grate_T0}) at zero temperature with the pre-quench parameter $h_0=0.5$. Nonanalyticity appears only at the phase boundary $h=0$. }
\label{T0_Results2}
\end{figure}

\begin{figure}
\includegraphics[scale=.16]{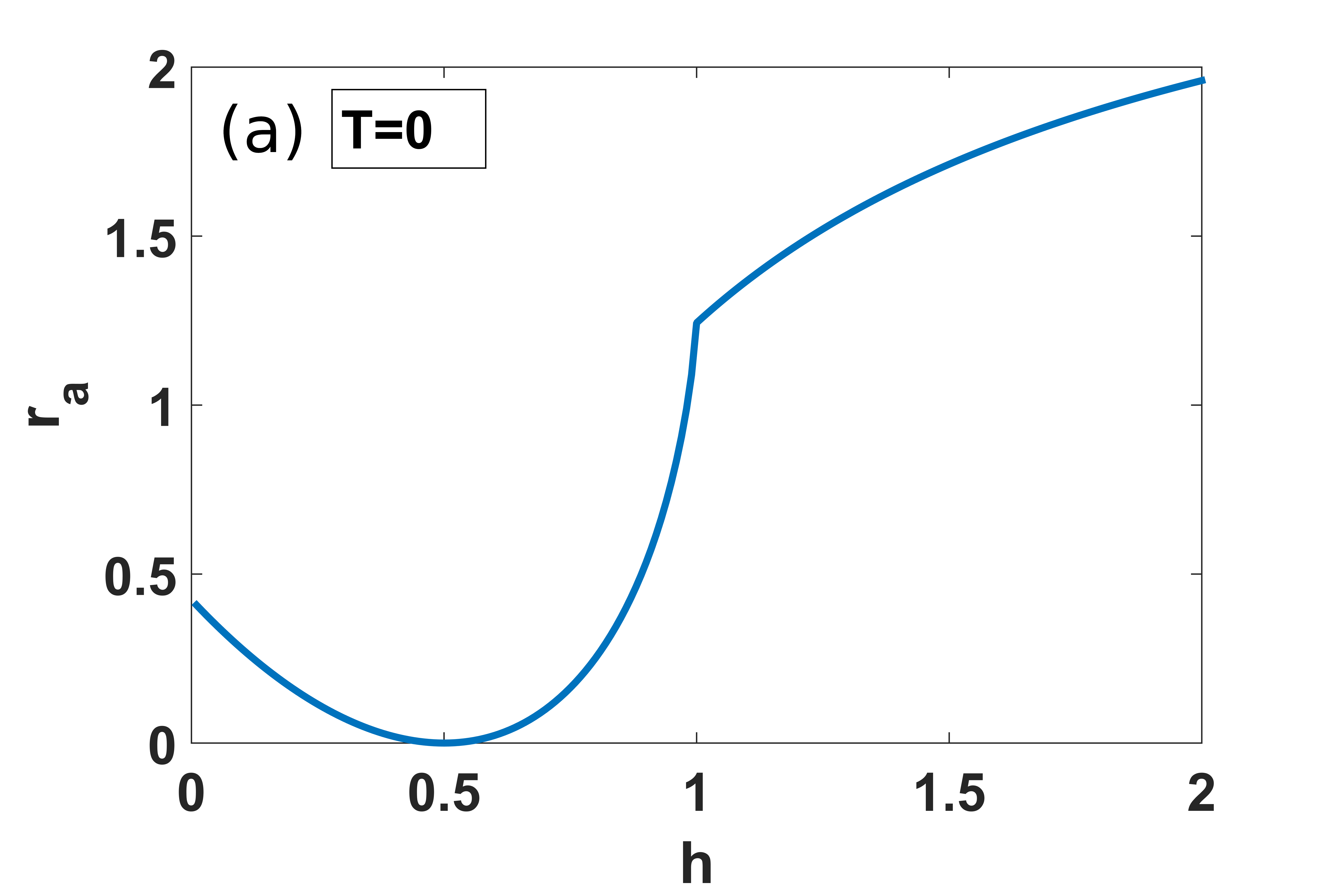}
\includegraphics[scale=.16]{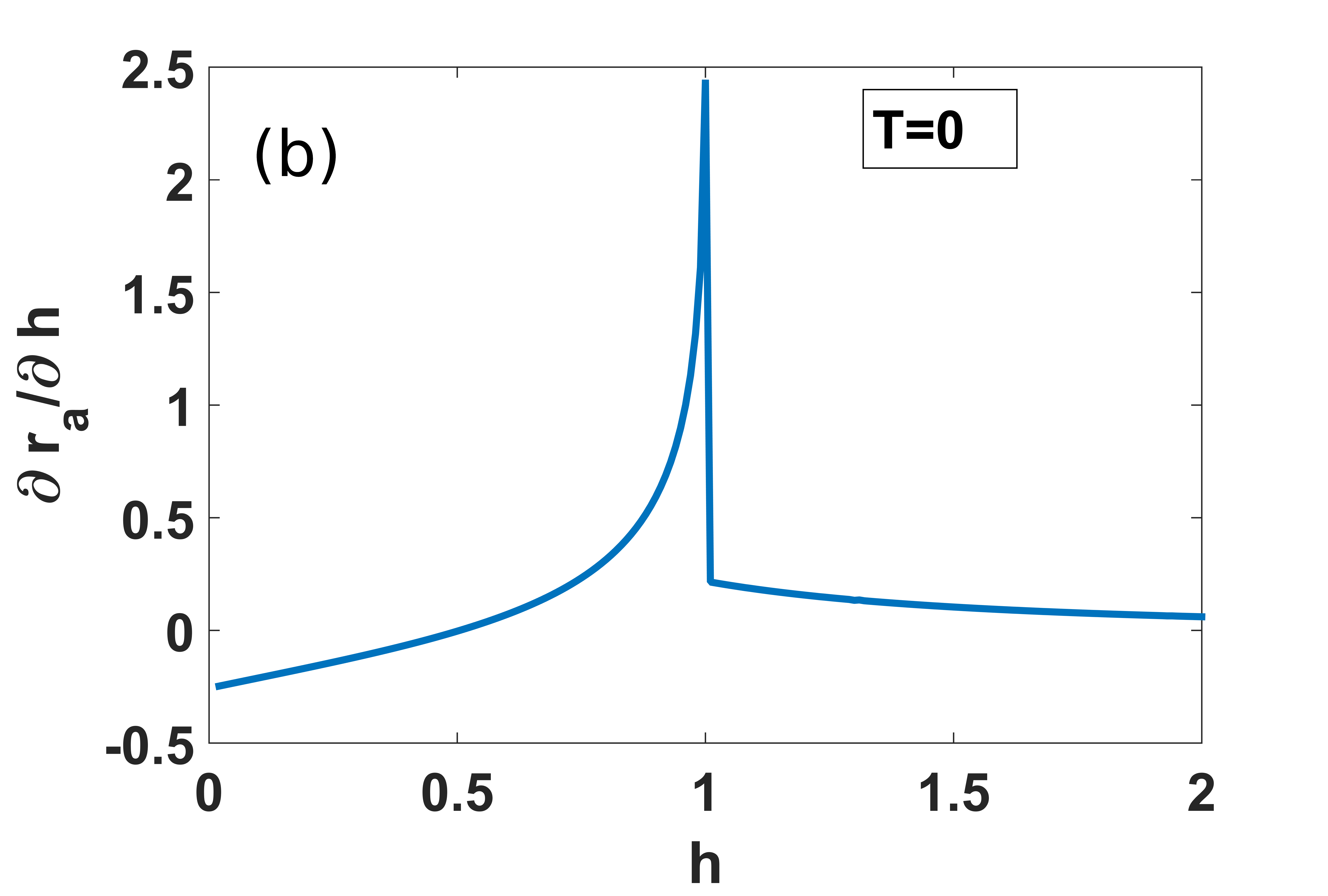}
\caption{Rate function $r_a$ (a) and its derivative (b) for 1D SSH Model computed from Eq. (\ref{grate_T0}) at zero temperature with the pre-quench parameter $h_0=0.5$. Nonanalyticity appears only at the phase boundary $h=1$. }
\label{T0_Results3}
\end{figure}

\begin{figure}
\includegraphics[scale=.31]{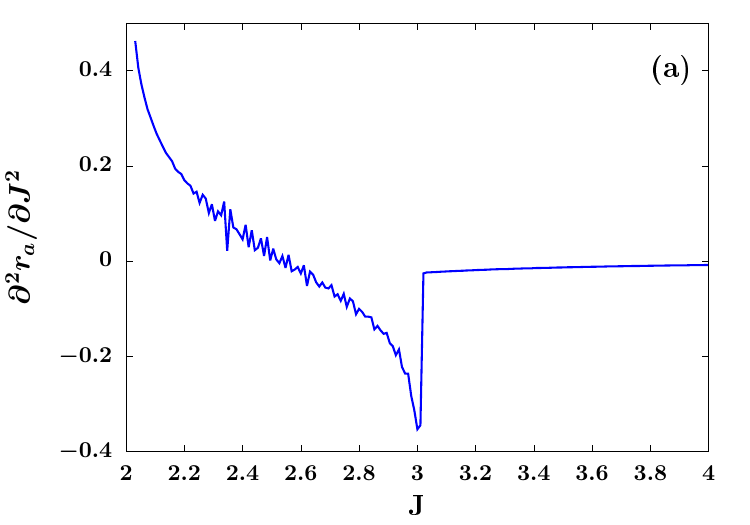}
\includegraphics[scale=.16]{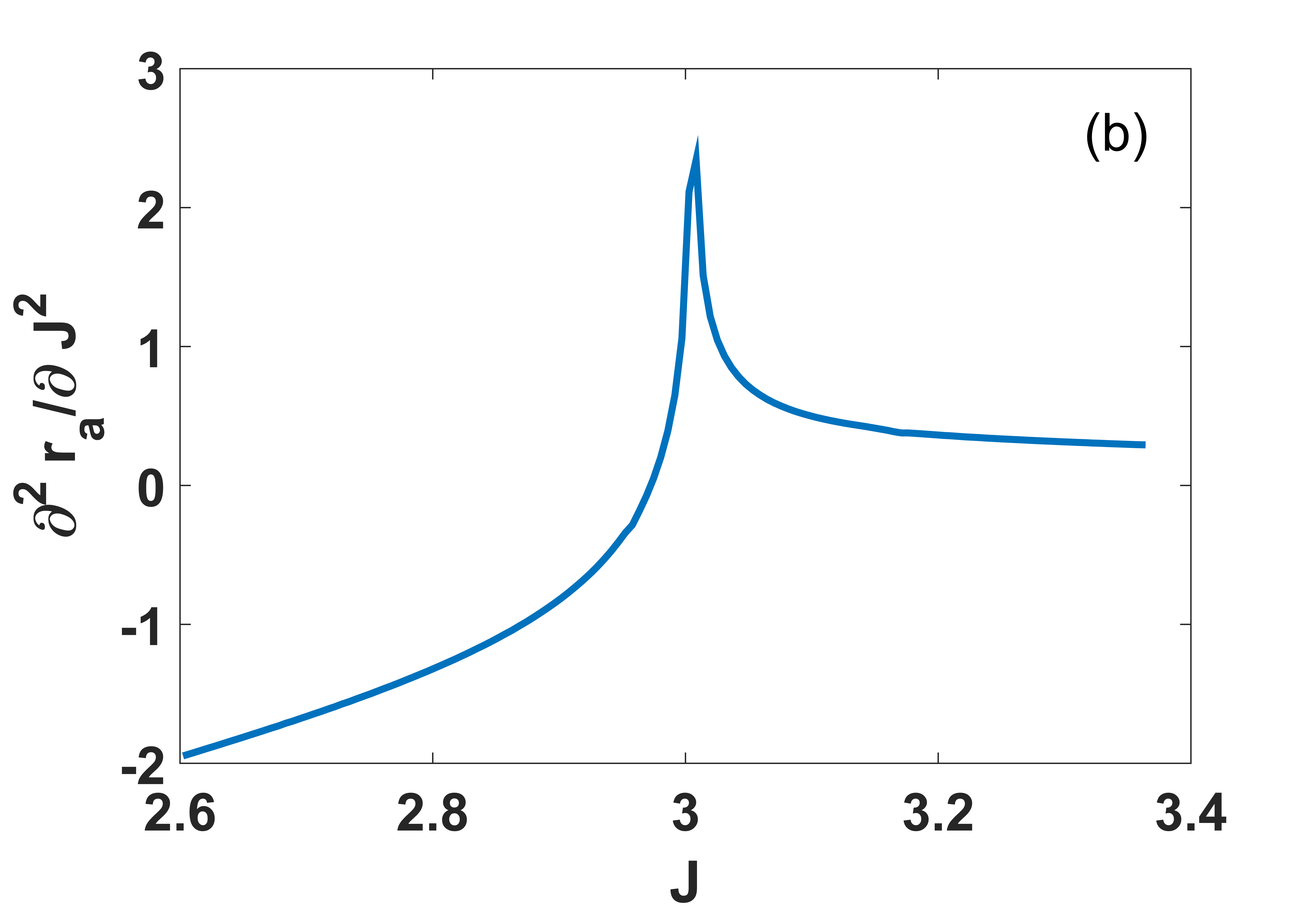}
\caption{Rate function $r_a$ and its derivative for (a) 3D topological nodal line semimetal and (b) Weyl Semimetal computed from Eq. (\ref{grate_T0}) at zero temperature with $J_0=2$ for TNLSM and $J_0=2.5$ for Weyl semimetal . Nonanalyticity appears only at the phase boundary $J=3$. The oscillatons in plot (a) is due to computational constraint.}
\label{T0_Results4}
\end{figure}

We have plotted this function and its derivative for diferent cases 
in Fig. (\ref{T0_Results1}),(\ref{T0_Results2}),(\ref{T0_Results3}),(\ref{T0_Results4}). The plots shows nonanalyticity only at the the quantum critical point.

\section{Conclusion}\label{conclu}
We explore the response of the fidelity at finite temperature in topological systems which can be mapped to free fermionic Hamiltonians and observe a nonanalyticity at the corresponding phase boundaries in different dimensions. The rate function in this case can be written 
as a series, each term of which has an integral independent of temperature, while the pre-factors of the integrals contain temperature. We could show that the integral brings about the nonanalytic behavior and hence the signature persists at high temperature also.

\par In case of 1D and 2D Hamiltonians, the reason behind the nonanalytic behavior of the quantity could be explained by mapping the integral to the complex plane and identifying a change of pole structure at the quantum critical point. The amount of the discontinuity or the exponent of the divegence as the case may be, could also be determined. However, analytical explanation was not possible in the case of 3D Hamiltonians. We suspect that the numerical limitation is the reason behind the signature being visible at low temperatures only. More investigation in this direction is in progress.
\par We have also shown that the long time limit of the rate function $r_a$ shows nonanalyticity only at QCP at $T=0$ while the nonanalyticity in rate funcion for Loschmidt echo vs time plots may not correspond with a quench across QCP.
\par A question that immediately comes to mind is whether this signature has the robustness beyond 3-dimension. Our primary conjecture is that it has the robustness because the concerned integral would just turn into a $d$-dimensional one. However more detailed studies are needed to confirm this. It would also be interesting to explore how our rate function behaves for other integrable and nonintegrable Hamiltonians.


\section*{Acknowledgements}
\noindent
The authors acknowledge anonymous referees for constructive opinions on an early version of this work. PN acknowledges UGC for financial support (Ref. No. 191620072523) and Harish Chandra Research Institute for access to their infrastructure.

\section*{Appendix A}
\renewcommand{\theequation}{A.\arabic{equation}}
\setcounter{equation}{0}
\noindent
In this Appendix we present the detailed derivation for the expressions of rate function in Eqs. (\ref{prev_eqn8}), (\ref{rddim}) and (\ref{rlog1}). From the expressions for $\mathcal{H}_{\vec{q}}$ and $\mathcal{H}'_{\vec{q}}$ in Eq. (\ref{expression_Hq}) and the equation next to it, we can write the matrix product of $(\hat{\mathcal{V}}_{\vec{q}}\cdot \vec{\sigma})$ and $(\hat{\mathcal{V}}'_{\vec{q}}\cdot \vec{\sigma}) $ as
\[ (\hat{\mathcal{V}}_{\vec{q}}\cdot \vec{\sigma})(\hat{\mathcal{V}}'_{\vec{q}}\cdot \vec{\sigma})  = 
\vec{{\mathcal V}} \cdot \vec{{\mathcal V}'}  \; \mbox{\underline{1}} + i \left[\vec{{\mathcal V}} \times \vec{{\mathcal V}'} \right] \cdot \vec{\sigma}\]
Since
\[ \mathcal{H}_{\vec{q}}^2=\lambda_{\vec{q}}^2 \underline{1} \;\;\;
\mbox{and} \;\;\; {\mathcal{H}'_{\vec{q}}}^2={\lambda'_{\vec{q}}}^2 \underline{1}\]
we get
\bea && e^{-\beta {\mathcal H}_{\vec{q}}} \cdot e^{-it{\mathcal H}'_{\vec{q}}}    \nonumber \\
 & = & \left[ \cosh (\beta \lambda) \cos (\lambda't) + i \sinh (\beta \lambda) \sin (\lambda't) \, \vec{{\mathcal V}} \cdot \vec{{\mathcal V}'}  \right] \mbox{\underline{1}}  - \nonumber \\
 && \vec{\sigma}\cdot  \left[ \sinh (\beta \lambda) \cos (\lambda't) \,\vec{\mathcal V} + i\cosh (\beta \lambda) \sin (\lambda't) \,  \vec{\mathcal V}' +\right. \nonumber \\
&& \left. \sinh (\beta \lambda) \sin (\lambda't) \, \vec{\mathcal V} \times \vec{\mathcal V}' \right]
 \eea
By reversing the sign of $t$ in this expression, one can obtain the expression for $ \rho_0e^{it{\mathcal H}'}$. One can now calculate $\rho_0 \rho_t = \rho_0e^{-it{\mathcal H}'} \rho_0e^{it{\mathcal H}'} $ and take the trace of it. The result is
 \be \frac{{\rm Tr} (\rho_0 \rho_t)}{[{\rm Tr} (\rho_0)]^2} = \frac{1}{2} \left[ 1 + \tanh^2 (\beta \lambda) - 2 {\mathcal L} \,\tanh^2 (\beta \lambda) \sin^2 (\lambda' t) \right] \ee
 where
\be {\mathcal L} = \frac{1}{2} \left[1 + \left|\vec{\mathcal V} \times \vec{\mathcal V}' \right|^2 - \left(\vec{\mathcal V} \cdot  \vec{\mathcal V}' \right)^2 \right] \ee
Since $\vec{\mathcal V}$ and $\vec{\mathcal V}'$ are unit vectors,
\be {\mathcal L} = 1 -   \left|\vec{\mathcal V} \cdot  \vec{\mathcal V}' \right|^2 = \sin^2 (\phi) \ee
where $\phi$ is the angle between the vectors $\vec{\mathcal V}$ and $\vec{\mathcal V}'$. \\

Using standard results \citep{gradshteyn2014}, we get the long-time limit as,
\bea r_a(\beta,p_0,p) &=& 3\log 2-\frac{1}{V}\int_{\vec{q}} \log(1+\alpha_{\vec{q}}) d\vec{q}\nonumber \\ &-&\frac{2}{V} \int_{\vec{q}} \log \left[1+\sqrt{1-\gamma_{\vec{q}} \mathcal{L}_{\vec{q}}}\right] d\vec{q}  \eea 
with,
 $\alpha_{\vec{q}}=\tanh^2(\beta \lambda_{\vec{q}})$, $\gamma_{\vec{q}}=2\alpha_{\vec{q}}/(1+\alpha_{\vec{q}})=1 - \text{sech} (2\beta \lambda_{\vec{q}})$ and $V=\int_{\vec{q}} d\vec{q}$.
We now note that $0 <\gamma_{\vec{q}}\mathcal{L}_{\vec{q}} < 1$ and use the expansion
\be \log \left[ 1+ \sqrt{1-x} \right] = \log 2+ \frac{1}{4} \left( x + \frac{3}{8}x^2 +\frac{5}{24}x^3 \cdots \right) \ee
to arrive at Eq. (\ref{rlog1}).
\label{appendixa}

\section*{Appendix B}
\renewcommand{\theequation}{B.\arabic{equation}}
\setcounter{equation}{0}
\noindent
In this Appendix, we consider Kitaev model on honeycomb lattice and show analytically that the double derivative diverges at the critical point with an exponent $1/2$ at all temperatures.
We consider the quench of $J_3$ from $J_0$ to $J$ at time $t=0$.  

We shall start by calculating $I_1$ which we obtain by putting $n=1$ in Eq.(\ref{In})

\bea I_1 =  4(J-J_0)^{2} \int_{u=0}^{\pi/2} du \; \int_{v=-\pi}^{\pi} dv   \;    \left(\frac{a'}{\lambda'}\right)^{2} \label{I1} \eea

where the primed quantities are the value of $b$ and $\lambda$ after quench. We have dropped the subscript $\v$ for brevity.

By substituting $z=e^{iv}$, we can write $I_1$ as 
\bea I_1 &=&-\frac{i(J-J_0)^2}{J}\int_0^{\pi \over 2} du\;\; \oint_{\mathcal{C}} dz F(z)  \eea

where $\mathcal{C}$ is the unit circle and 
\bea
F(z)=\frac{z_0(z^2-{\bar{z_0}\over z_0})^2}{z^2 (z-{J\over z_0})(z-{\bar{z_0}\over J})}
\eea
with $z_0=M\cos (u) +iN\sin (u)$

The poles of $F(z)$ are $z_1=0$, $z_2 ={J}/{z_0}$ and $z_3 = {\bar{z_0}}/{J}$ and the respctive residues are 
\bea R_1 &=& {\bar{z_0}\over J}+{J\over z_0} \no \\
 R_2 &=& -{\bar{z_0}\over J}+{J\over z_0} \no \\
 R_3 &=& {\bar{z_0}\over J}-{J\over z_0}  \eea
The poles $z_1$ and $z_2$ are inside the unit circle when $u<u_c$ and $z_1$ and $z_3$ are inside the unit circle when $u>u_c$ where $u_c$ follows the relation
\bea \cos (u_c)=\sqrt{\frac{(J+N)(J-N)}{4J_1 J_2}} \no  \eea
By applying the residue theorem, we can evaluate the integral $I_1$ as
\bea 
I_1 = {2\pi \over J}(J-J_0)^2\left[2J u_c-(M^2-N^2)\frac{\sin (2u_c)}{2J} +\right. \no \\
\left. {1 \over J}(M^2+N^2)({\pi \over 2}-u_c) \right] \no \\
\label{I1_final_result} 
\eea

When $J$ approaches $M$ from below, we can replace $J$ by $M-\epsilon$ in Eq. (\ref{I1_final_result}) where $\epsilon$ is very small and differentiating $I_1$ with respect to $J$ we get

\bea \frac{\partial ^2 I_1}{\partial J^2} \approx -\frac{4\pi\sqrt{2}(M-J_0)^2}{(M(M^2-N^2))^{1\over 2}} \epsilon^{-{1\over 2}}   \eea  
upto leading order.

When $J$ approaches $|N|$ from below, we can replace $J$ by $|N|+\epsilon$ in Eq. (\ref{I1_final_result}) where $\epsilon$ is small and differentiating $I_1$ with respect to $J$ we get
\bea \frac{\partial ^2 I_1}{\partial J^2} \approx -\frac{4\pi\sqrt{2}(|N|-J_0)^2}{(|N|(M^2-N^2))^{1\over 2}} \epsilon^{-{1\over 2}}   \eea
upto leading order.

This proves that at high temperature where $I_1$ is the dominant term, the double derivative of the rate function will diverge with an exponent $1/2$ at both the critical points. But at lower temperature, $I_n$ with $n>1$ will also contribute to the double derivative. We will show that the critical exponent will be unchanged for $I_n$ with $n>1$. We need to calculate the integral $I_n$, which after substitution $z=e^{iv}$ can be written as 
\bea I_n = -\frac{4i(J-J_0)^{2n}}{(4J)^{n}} \int_{0}^{\pi \over 2} du \oint F^{(n)} dz   \eea
where 
\bea F^{(n)} = \frac{z_0^n (z^2-\frac{\bar{z_0}}{z_0})^{2n}}{z^{n+1}(z-\frac{J}{z_0})^n(z-\frac{\bar{z_0}}{J})^n}  \eea
By applying residue theorem, we can write the integral $I_n$ as 
\bea I_n = \frac{8\pi(J-J_0)^{2n}}{(4J)^{n}}\left[\int_0^{\frac{\pi}{2}} R_1^{(n)} du + \right. \no \\
\left. \int^{u_c}_0 R_2^{(n)} du + \int^{\frac{\pi}{2}}_{u_c} R_3^{(n)} du\right] \no \eea
where $R_1^{(n)}$, $R_2^{(n)}$ and $R_3^{(n)}$ are the residues of the poles $z_1=0$, $z_2={J}/{z_0}$ and $z_3={\bar{z_0}}/{J}$ respectively. 

Since the nonanalyticity arises from the residues $R_2^{(n)}$ and $R_3^{(n)}$, we will prove that both $R_2^{(n)}$ and $R_3^{(n)}$ are proportional to $R_2$ and $R_3$ respectively upto leading order.

We can write $R_2^n$ as 
\bea R_2^{(n)} &=& \lim_{z\rightarrow z_2}\; \frac{1}{(n-1)!} \;\frac{d^{n-1}}{dz^{n-1}}\; \frac{z_0^n (z^2-\frac{\bar{z_0}}{z_0})^{2n}}{z^{n+1}(z-\frac{\bar{z_0}}{J})^n} \no \\
&=& \lim_{z\rightarrow z_2}\; \frac{1}{(n-1)!} \;\frac{d^{n-1}}{dz^{n-1}}\; \frac{G^n}{z}  \eea
where we define $G(z)$ as 
\bea G(z) = \frac{z_0(z^2-\frac{\bar{z_0}}{z_0})^2}{z(z-\frac{\bar{z_0}}{J})} \eea
We observe that 
\bea \frac{d^{n-1}(G^n/z)}{dz^{n-1}} = \frac{G}{z}\left[n! \left(\frac{dG}{dz}\right)^{n-1}+\Gamma \right] \eea
where $\Gamma$ contains the terms which will involve $G^k$ or $G^k/z$ as a factor where $k \geqslant 1$.
We know that $\left({G}/{z}\right)_{z=z_2}=R_2^1$ and $\left({dG}/{dz}\right)_{z=z_2} = 2J+{z_0\bar{z_0}}/{J}$.
Therefore
\bea \left(\frac{dG}{dz} |_{z=z_2}\right)^{n-1} \approx \left(\frac{2M^2+z_0\bar{z_0}}{M}\right)^{n-1}(1+ c\epsilon)^{n-1} \no \\ \;\;\mbox{when}\; J=M-\epsilon \no \label{dgdzn-1M}\\
\left(\frac{dG}{dz} |_{z=z_2}\right)^{n-1} \approx \left(\frac{2N^2+z_0\bar{z_0}}{|N|}\right)^{n-1}(1+ c'\epsilon)^{n-1}\no \\ \;\;\mbox{when}\; J=|N|+\epsilon \no \label{dgdzn-1N} \eea
where $c$ and $c'$ are constants.
Putting the values of $\left(\frac{dG}{dz} |_{z=z_2}\right)^{n-1}$ and $\left(\frac{G}{z}\right)_{z=z_2}$, we get 
\bea R_2^{(n)} \approx n \left(\frac{2M^2+z_0\bar{z_0}}{M}\right)^{n-1} R_2^{(1)} \;\;\mbox{when}\; J=M-\epsilon \no \\
R_2^{(n)}\approx n \left(\frac{2N^2+z_0\bar{z_0}}{|N|}\right)^{n-1} R_2^{(1)}\;\; \mbox{when}\; J=|N|+\epsilon \no \\
\eea 
Similar calculation can be done for $R_3^{(n)}$.
Thus we prove that $I_n$ will also diverge with an exponent $1/2$ at both quantum critical points.

\label{appendixb}

\section*{Appendix C}
\renewcommand{\thefigure}{C.\arabic{figure}}
\setcounter{figure}{0}
\noindent
In the figures \ref{timodelI1},\ref{xymodelI1} and \ref{sshmodelI1} we show numerically that the discontinuity in the single derivative of the rate function with respect to the final value of the quench parameter can be approximated by the derivative of the $n=1$ term in Eq (\ref{rlog}) for the one dimensional models at high temperatures. The fact that near critical point behaviour of the rate function can be approximated by $I_1$ even for 2D Kitaev Model at high temperatures was established in \cite{nandi2022}.   

\begin{figure}[htbp]
    \includegraphics[scale=0.32]{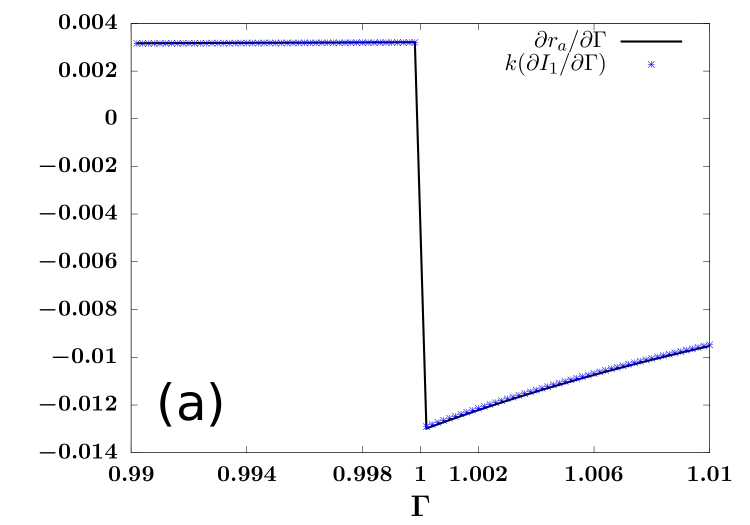}
    \includegraphics[scale=0.32]{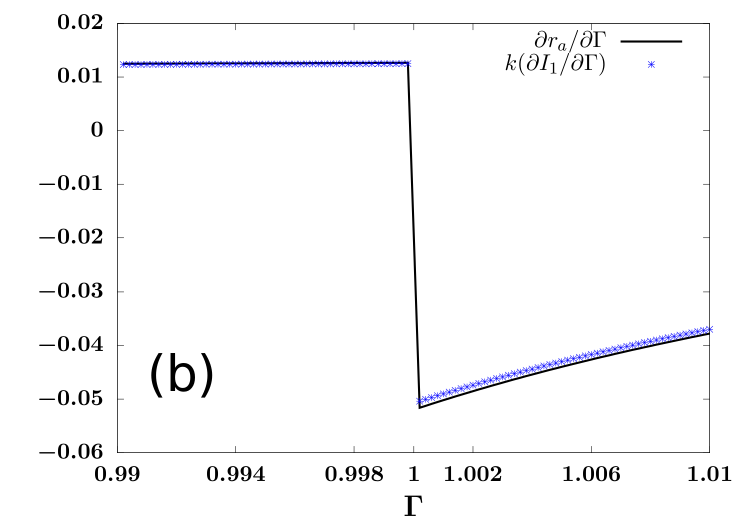}
    \caption{Transverse XY Model : The plot shows that discontinuity in $\partial r_a/\partial \Gamma$ comes mainly from the discontinuity in $\partial I_1/\partial \Gamma$ at high temperatures. $k$ is the constant $(c_1 \gamma)/(2V\lambda^2)$ which is multiplied to $I_1$ in Eq (\ref{rlog}). The figure (a) is for $\beta=0.1$ and  figure (b) is for $\beta=0.2$. Both the cases are for quench of $\Gamma$ parameter.}
    \label{timodelI1}
\end{figure}

\begin{figure}[htbp]
    \includegraphics[scale=0.32]{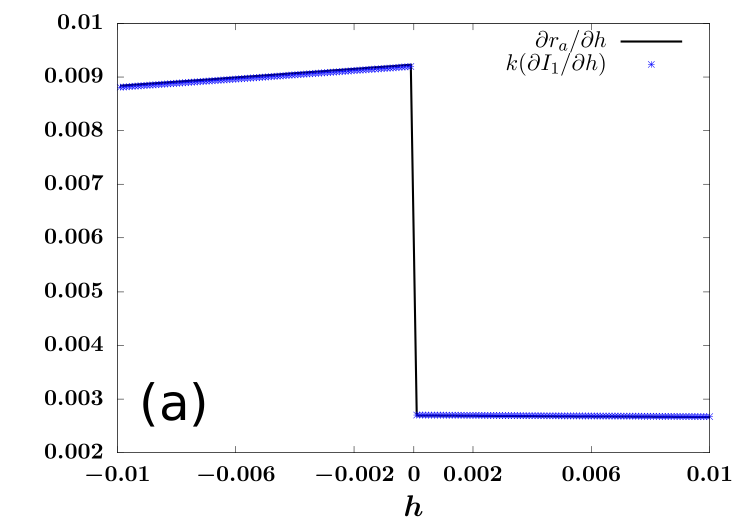}
    \includegraphics[scale=0.32]{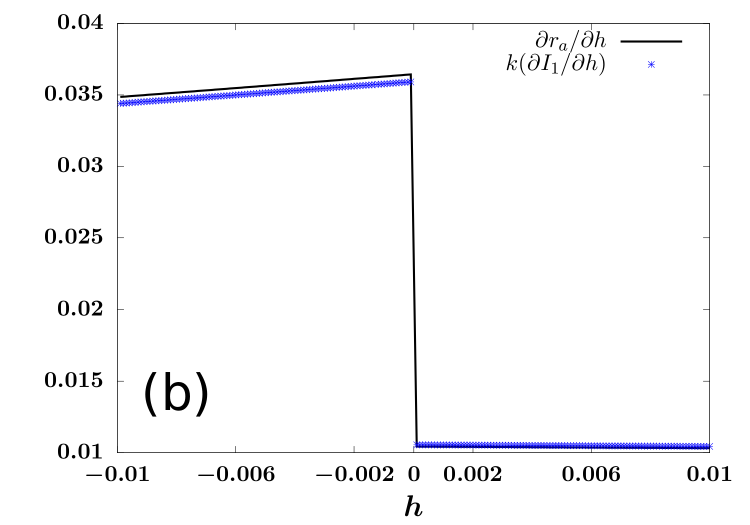}
    \caption{Transverse XY Model : The plot shows that discontinuity in $\partial r_a/\partial h$ comes mainly from the discontinuity in $\partial I_1/\partial h$ at high temperatures. $k$ is the constant $(c_1 \gamma)/(2V\lambda^2)$ which is multiplied to $I_1$ in Eq (\ref{rlog}). The figure (a) is for $\beta=0.1$ and  figure (b) is for $\beta=0.2$. Both the cases are for quench of $h$ parameter.}
    \label{xymodelI1}
\end{figure}
\begin{figure}[htbp]
    \includegraphics[scale=0.32]{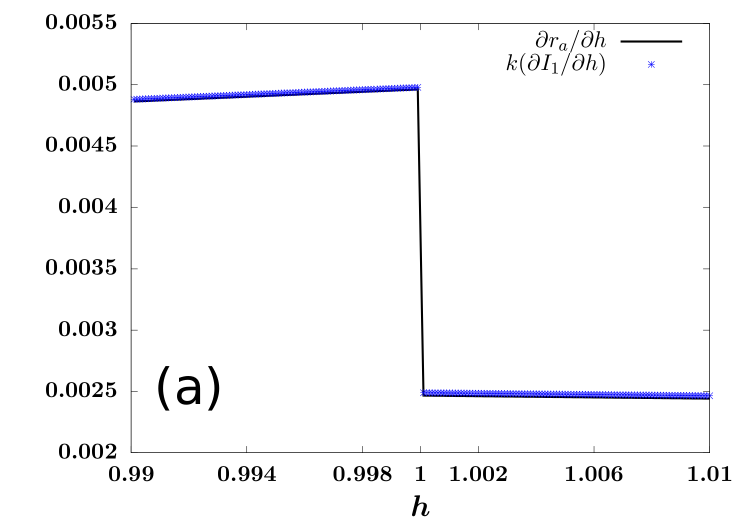}
    \includegraphics[scale=0.32]{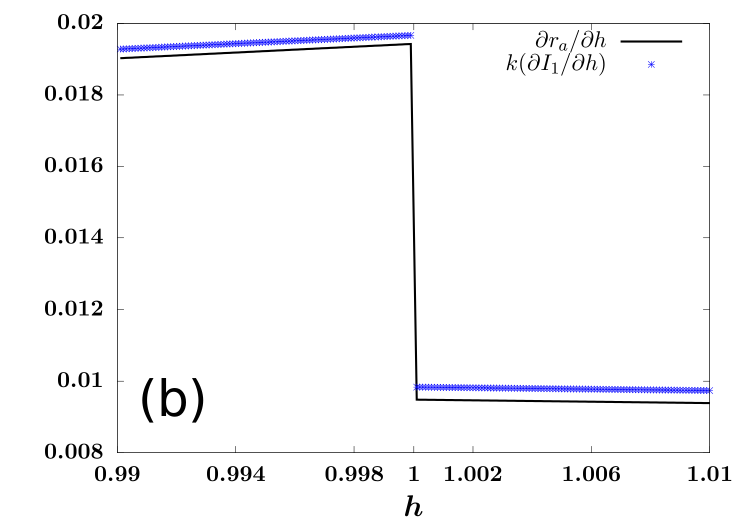}
    \caption{SSH Model :The plot shows that discontinuity in $\partial r_a/\partial h$ comes mainly from the discontinuity in $\partial I_1/\partial h$ at high temperatures. $k$ is the constant $(c_1 \gamma)/(2V\lambda^2)$ which is multiplied to $I_1$ in Eq (\ref{rlog}). The figure (a) is for $\beta=0.1$ and  figure (b) is for $\beta=0.2$. Both the cases are for quench of $h$ parameter.}
    \label{sshmodelI1}
\end{figure}
\label{appendixc}
\clearpage
\bibliography{dmt}

\end{document}